\documentstyle[10pt,epsf,epsfig,hangcaption,xspace,amssymb,amsfonts,amsmath,amsthm,cite,dp_delphititle,lineno,subfigure]{dp_delphi}
\makeindex
\def\DpTitle{{
Searches for invisibly decaying Higgs bosons with the DELPHI detector at LEP}
}
\def\DpPaperGroup{EP}
\def\DpPaperRef{2003-046}
\def\DpDate{18 June 2003}
\def\DpAuthors{DELPHI Collaboration}
\def\DpSubmit{(Accepted by Eur. Phys. J. C)}
\def\DpComment{ }
\def\DpEMail{ }
\def\Hz{\mbox{H}}
\def\hnn{\mbox{\Hz$\mathrm \nu \bar{\nu} $}}

\newcommand{\hmumu}{\mbox{${\mathrm{H\mu^+\mu^-}}$}}
\newcommand{\hee}{\mbox{${\mathrm{He^+e^-}}$}}
\newcommand{\htautau}{\mbox{${\mathrm{H\tau^+\tau^-}}$}}
\newcommand{\hqq}{\mbox{${\mathrm{Hq\bar{q}}}$}}

\newcommand{\ee} {\mbox{$ {\mathrm e}^+ {\mathrm e}^- $}}
\newcommand{\GeV} {\mbox{$ {\mathrm{GeV}} $}}
\newcommand{\GeVcc} {\mbox{$ {\mathrm{GeV}}/\mathrm{c^2} $}}
\newcommand{\ffbar} {\mbox{$ {\mathrm q}\bar{\mathrm q} $}}
\newcommand{\qqev} {\mbox{$ \rm q\bar{q} e \bar{\nu} $}}
\newcommand{\EGSSS}
{\mbox{${E_{\mathrm \gamma}/E^{\rm Z}_{\mathrm \gamma}} $}}
\newcommand{\MZ} {\mbox{$ {m}_{{\mathrm Z}} $}}
\newcommand{\Evis} {\mbox{$ {E}_{\mathrm{vis}} $}}
\newcommand{\Mvis} {\mbox{$ {m}_{\mathrm{vis}} $}}
\newcommand{\GeVc} {\mbox{$ {\mathrm{GeV}}/c $}}
\def\pbinv {\mbox{pb$^{-1}$}}
\def\NPB#1#2#3{{\it Nucl.~Phys.} {\bf{B#1}} (19#2) #3}
\def\PLB#1#2#3{{\it Phys.~Lett.} {\bf{B#1}} (19#2) #3}
\def\PLBN#1#2#3{{\it Phys.~Lett.} {\bf{B#1}} (20#2) #3}
\def\PRD#1#2#3{{\it Phys.~Rev.} {\bf{D#1}} (19#2) #3}

\def\ZPC#1#2#3{{\it Z.~Phys.} {\bf C#1} (19#2) #3}

\def\NIMA#1#2#3{{\it Nucl.~Instr.~and~Meth.} {\bf#1} (19#2) #3}
\def\CPC#1#2#3{{\it Comp.~Phys.~Comm.} {\bf#1} (19#2) #3}
\def\CPCY#1#2#3{{\it Comp.~Phys.~Comm.} {\bf#1} (20#2) #3}
\def\EPJ#1#2#3{{\it Eur.~Phys.~J.} {\bf#1} (#2) #3}
\begin{document}
%%%%%%%%%%%%%%%%%%%%%%%%%%%%
%AS: add EPS cover page:
%
%\includegraphics[width=\textwidth]{inv03.ps}
%\cleardoublepage
%
%%%%%%%%%%%%%%%%%%%%%%%%%%%%
%%%%%%%%%%%%%%%%%%%%%%%%%% They are a problem with Coll.Sty ?
\makeatletter
%\input{dp_system:coll.sty}
% Collapse citation numbers to ranges.  Non-numeric and undefined labels
% are handled.  No sorting is done.  E.g., 1,3,2,3,4,5,foo,1,2,3,?,4,5
% gives 1,3,2-5,foo,1-3,?,4,5
\newcount\@tempcntc
\def\@citex[#1]#2{\if@filesw\immediate\write\@auxout{\string\citation{#2}}\fi
  \@tempcnta\z@\@tempcntb\m@ne\def\@citea{}\@cite{\@for\@citeb:=#2\do
    {\@ifundefined
       {b@\@citeb}{\@citeo\@tempcntb\m@ne\@citea\def\@citea{,}{\bf ?}\@warning
       {Citation `\@citeb' on page \thepage \space undefined}}%
    {\setbox\z@\hbox{\global\@tempcntc0\csname b@\@citeb\endcsname\relax}%
     \ifnum\@tempcntc=\z@ \@citeo\@tempcntb\m@ne
       \@citea\def\@citea{,}\hbox{\csname b@\@citeb\endcsname}%
     \else
      \advance\@tempcntb\@ne
      \ifnum\@tempcntb=\@tempcntc
      \else\advance\@tempcntb\m@ne\@citeo
      \@tempcnta\@tempcntc\@tempcntb\@tempcntc\fi\fi}}\@citeo}{#1}}
\def\@citeo{\ifnum\@tempcnta>\@tempcntb\else\@citea\def\@citea{,}%
  \ifnum\@tempcnta=\@tempcntb\the\@tempcnta\else
   {\advance\@tempcnta\@ne\ifnum\@tempcnta=\@tempcntb \else \def\@citea{--}\fi
    \advance\@tempcnta\m@ne\the\@tempcnta\@citea\the\@tempcntb}\fi\fi}
 
\makeatother
%%%%%%%%%%%%%%%%%%%%%%%%%% ??????????????????????????????????
% Generate the title page
\begin{titlepage}
\pagenumbering{roman}
\CERNpreprint{\DpPaperGroup}{\DpPaperRef} % Reference of the paper
\date{{\small\DpDate}} % Date of the paper
\title{\DpTitle} % Title of the paper
\address{\DpAuthors} % General name of the author(s)
\begin{shortabs} % Start the abstract
\noindent
\noindent
Searches for {\mbox{$ {\mathrm H} {\mathrm Z}$}} production
with the Higgs boson decaying into an invisible final state
were performed using the data collected by the DELPHI
experiment at centre-of-mass energies between 188 GeV and 209 GeV.
Both hadronic and leptonic final states of the Z boson were analysed.
In addition to the search for a heavy Higgs boson, a dedicated search
for a light Higgs boson down to 40~\GeVcc~was performed.
No signal was found.
Assuming the Standard Model HZ production
cross-section, the mass limit for invisibly decaying Higgs 
bosons is 112.1~\GeVcc~at 95\% confidence level.
An interpretation in the Minimal Supersymmetric extension of the Standard
Model (MSSM) and in a Majoron model is also given.

%=========================================================================%
\end{shortabs}
\vfill
\begin{center}
\DpSubmit \ \\ % Horrible hack to allow to have DpSubmit empty
\DpComment \ \\
\DpEMail \ \\
\end{center}
\vfill
\clearpage
\headsep 10.0pt
\addtolength{\textheight}{10mm}
\addtolength{\footskip}{-5mm}
\begingroup
% Commands to process the author names
%
\newcommand{\DpName}[2]{\hbox{#1$^{\ref{#2}}$},\hfill}
\newcommand{\DpNameTwo}[3]{\hbox{#1$^{\ref{#2},\ref{#3}}$},\hfill}
\newcommand{\DpNameThree}[4]{\hbox{#1$^{\ref{#2},\ref{#3},\ref{#4}}$},\hfill}
\newskip\Bigfill \Bigfill = 0pt plus 1000fill
\newcommand{\DpNameLast}[2]{\hbox{#1$^{\ref{#2}}$}\hspace{\Bigfill}}
%
%\small
\footnotesize
\noindent
\DpName{J.Abdallah}{LPNHE}
\DpName{P.Abreu}{LIP}
\DpName{W.Adam}{VIENNA}
\DpName{P.Adzic}{DEMOKRITOS}
\DpName{T.Albrecht}{KARLSRUHE}
\DpName{T.Alderweireld}{AIM}
\DpName{R.Alemany-Fernandez}{CERN}
\DpName{T.Allmendinger}{KARLSRUHE}
\DpName{P.P.Allport}{LIVERPOOL}
\DpName{U.Amaldi}{MILANO2}
% ...
\normalsize
\endgroup
\titlefoot{Department of Physics and Astronomy, Iowa State
     University, Ames IA 50011-3160, USA
    \label{AMES}}
\titlefoot{Physics Department, Universiteit Antwerpen,
     Universiteitsplein 1, B-2610 Antwerpen, Belgium \\
     \indent~~and IIHE, ULB-VUB,
     Pleinlaan 2, B-1050 Brussels, Belgium \\
     \indent~~and Facult\'e des Sciences,
     Univ. de l'Etat Mons, Av. Maistriau 19, B-7000 Mons, Belgium
    \label{AIM}}
\titlefoot{Physics Laboratory, University of Athens, Solonos Str.
     104, GR-10680 Athens, Greece
    \label{ATHENS}}
\titlefoot{Department of Physics, University of Bergen,
     All\'egaten 55, NO-5007 Bergen, Norway
    \label{BERGEN}}
\titlefoot{Dipartimento di Fisica, Universit\`a di Bologna and INFN,
     Via Irnerio 46, IT-40126 Bologna, Italy
    \label{BOLOGNA}}
\titlefoot{Centro Brasileiro de Pesquisas F\'{\i}sicas, rua Xavier Sigaud 150,
     BR-22290 Rio de Janeiro, Brazil \\
     \indent~~and Depto. de F\'{\i}sica, Pont. Univ. Cat\'olica,
     C.P. 38071 BR-22453 Rio de Janeiro, Brazil \\
     \indent~~and Inst. de F\'{\i}sica, Univ. Estadual do Rio de Janeiro,
     rua S\~{a}o Francisco Xavier 524, Rio de Janeiro, Brazil
    \label{BRASIL}}
\titlefoot{Coll\`ege de France, Lab. de Physique Corpusculaire, IN2P3-CNRS,
     FR-75231 Paris Cedex 05, France
    \label{CDF}}
\titlefoot{CERN, CH-1211 Geneva 23, Switzerland
    \label{CERN}}
\titlefoot{Institut de Recherches Subatomiques, IN2P3 - CNRS/ULP - BP20,
     FR-67037 Strasbourg Cedex, France
    \label{CRN}}
\titlefoot{Now at DESY-Zeuthen, Platanenallee 6, D-15735 Zeuthen, Germany
    \label{DESY}}
\titlefoot{Institute of Nuclear Physics, N.C.S.R. Demokritos,
     P.O. Box 60228, GR-15310 Athens, Greece
    \label{DEMOKRITOS}}
\titlefoot{FZU, Inst. of Phys. of the C.A.S. High Energy Physics Division,
     Na Slovance 2, CZ-180 40, Praha 8, Czech Republic
    \label{FZU}}
\titlefoot{Dipartimento di Fisica, Universit\`a di Genova and INFN,
     Via Dodecaneso 33, IT-16146 Genova, Italy
    \label{GENOVA}}
\titlefoot{Institut des Sciences Nucl\'eaires, IN2P3-CNRS, Universit\'e
     de Grenoble 1, FR-38026 Grenoble Cedex, France
    \label{GRENOBLE}}
\titlefoot{Helsinki Institute of Physics, P.O. Box 64,
     FIN-00014 University of Helsinki, Finland
    \label{HELSINKI}}
\titlefoot{Joint Institute for Nuclear Research, Dubna, Head Post
     Office, P.O. Box 79, RU-101 000 Moscow, Russian Federation
    \label{JINR}}
\titlefoot{Institut f\"ur Experimentelle Kernphysik,
     Universit\"at Karlsruhe, Postfach 6980, DE-76128 Karlsruhe,
     Germany
    \label{KARLSRUHE}}
\titlefoot{Institute of Nuclear Physics,Ul. Kawiory 26a,
     PL-30055 Krakow, Poland
    \label{KRAKOW1}}
\titlefoot{Faculty of Physics and Nuclear Techniques, University of Mining
     and Metallurgy, PL-30055 Krakow, Poland
    \label{KRAKOW2}}
\titlefoot{Universit\'e de Paris-Sud, Lab. de l'Acc\'el\'erateur
     Lin\'eaire, IN2P3-CNRS, B\^{a}t. 200, FR-91405 Orsay Cedex, France
    \label{LAL}}
\titlefoot{School of Physics and Chemistry, University of Lancaster,
     Lancaster LA1 4YB, UK
    \label{LANCASTER}}
\titlefoot{LIP, IST, FCUL - Av. Elias Garcia, 14-$1^{o}$,
     PT-1000 Lisboa Codex, Portugal
    \label{LIP}}
\titlefoot{Department of Physics, University of Liverpool, P.O.
     Box 147, Liverpool L69 3BX, UK
    \label{LIVERPOOL}}
\titlefoot{Dept. of Physics and Astronomy, Kelvin Building,
     University of Glasgow, Glasgow G12 8QQ
    \label{GLASGOW}}
\titlefoot{LPNHE, IN2P3-CNRS, Univ.~Paris VI et VII, Tour 33 (RdC),
     4 place Jussieu, FR-75252 Paris Cedex 05, France
    \label{LPNHE}}
\titlefoot{Department of Physics, University of Lund,
     S\"olvegatan 14, SE-223 63 Lund, Sweden
    \label{LUND}}
\titlefoot{Universit\'e Claude Bernard de Lyon, IPNL, IN2P3-CNRS,
     FR-69622 Villeurbanne Cedex, France
    \label{LYON}}
\titlefoot{Dipartimento di Fisica, Universit\`a di Milano and INFN-MILANO,
     Via Celoria 16, IT-20133 Milan, Italy
    \label{MILANO}}
\titlefoot{Dipartimento di Fisica, Univ. di Milano-Bicocca and
     INFN-MILANO, Piazza della Scienza 2, IT-20126 Milan, Italy
    \label{MILANO2}}
\titlefoot{IPNP of MFF, Charles Univ., Areal MFF,
     V Holesovickach 2, CZ-180 00, Praha 8, Czech Republic
    \label{NC}}
\titlefoot{NIKHEF, Postbus 41882, NL-1009 DB
     Amsterdam, The Netherlands
    \label{NIKHEF}}
\titlefoot{National Technical University, Physics Department,
     Zografou Campus, GR-15773 Athens, Greece
    \label{NTU-ATHENS}}
\titlefoot{Physics Department, University of Oslo, Blindern,
     NO-0316 Oslo, Norway
    \label{OSLO}}
\titlefoot{Dpto. Fisica, Univ. Oviedo, Avda. Calvo Sotelo
     s/n, ES-33007 Oviedo, Spain
    \label{OVIEDO}}
\titlefoot{Department of Physics, University of Oxford,
     Keble Road, Oxford OX1 3RH, UK
    \label{OXFORD}}
\titlefoot{Dipartimento di Fisica, Universit\`a di Padova and
     INFN, Via Marzolo 8, IT-35131 Padua, Italy
    \label{PADOVA}}
\titlefoot{Rutherford Appleton Laboratory, Chilton, Didcot
     OX11 OQX, UK
    \label{RAL}}
\titlefoot{Dipartimento di Fisica, Universit\`a di Roma II and
     INFN, Tor Vergata, IT-00173 Rome, Italy
    \label{ROMA2}}
\titlefoot{Dipartimento di Fisica, Universit\`a di Roma III and
     INFN, Via della Vasca Navale 84, IT-00146 Rome, Italy
    \label{ROMA3}}
\titlefoot{DAPNIA/Service de Physique des Particules,
     CEA-Saclay, FR-91191 Gif-sur-Yvette Cedex, France
    \label{SACLAY}}
\titlefoot{Instituto de Fisica de Cantabria (CSIC-UC), Avda.
     los Castros s/n, ES-39006 Santander, Spain
    \label{SANTANDER}}
\titlefoot{Inst. for High Energy Physics, Serpukov
     P.O. Box 35, Protvino, (Moscow Region), Russian Federation
    \label{SERPUKHOV}}
\titlefoot{J. Stefan Institute, Jamova 39, SI-1000 Ljubljana, Slovenia
     and Laboratory for Astroparticle Physics,\\
     \indent~~Nova Gorica Polytechnic, Kostanjeviska 16a, SI-5000 Nova Gorica, Slovenia, \\
     \indent~~and Department of Physics, University of Ljubljana,
     SI-1000 Ljubljana, Slovenia
    \label{SLOVENIJA}}
\titlefoot{Fysikum, Stockholm University,
     Box 6730, SE-113 85 Stockholm, Sweden
    \label{STOCKHOLM}}
\titlefoot{Dipartimento di Fisica Sperimentale, Universit\`a di
     Torino and INFN, Via P. Giuria 1, IT-10125 Turin, Italy
    \label{TORINO}}
\titlefoot{INFN,Sezione di Torino, and Dipartimento di Fisica Teorica,
     Universit\`a di Torino, Via P. Giuria 1,\\
     \indent~~IT-10125 Turin, Italy
    \label{TORINOTH}}
\titlefoot{Dipartimento di Fisica, Universit\`a di Trieste and
     INFN, Via A. Valerio 2, IT-34127 Trieste, Italy \\
     \indent~~and Istituto di Fisica, Universit\`a di Udine,
     IT-33100 Udine, Italy
    \label{TU}}
\titlefoot{Univ. Federal do Rio de Janeiro, C.P. 68528
     Cidade Univ., Ilha do Fund\~ao
     BR-21945-970 Rio de Janeiro, Brazil
    \label{UFRJ}}
\titlefoot{Department of Radiation Sciences, University of
     Uppsala, P.O. Box 535, SE-751 21 Uppsala, Sweden
    \label{UPPSALA}}
\titlefoot{IFIC, Valencia-CSIC, and D.F.A.M.N., U. de Valencia,
     Avda. Dr. Moliner 50, ES-46100 Burjassot (Valencia), Spain
    \label{VALENCIA}}
\titlefoot{Institut f\"ur Hochenergiephysik, \"Osterr. Akad.
     d. Wissensch., Nikolsdorfergasse 18, AT-1050 Vienna, Austria
    \label{VIENNA}}
\titlefoot{Inst. Nuclear Studies and University of Warsaw, Ul.
     Hoza 69, PL-00681 Warsaw, Poland
    \label{WARSZAWA}}
\titlefoot{Fachbereich Physik, University of Wuppertal, Postfach
     100 127, DE-42097 Wuppertal, Germany
    \label{WUPPERTAL}}
\addtolength{\textheight}{-10mm}
\addtolength{\footskip}{5mm}
\clearpage
\headsep 30.0pt
\end{titlepage}
%%%%%%%%%%%%%%%%%%%%%%%%%
%
% Change for the document body
%%\pagestyle{heading} % for page numbering
\pagenumbering{arabic} % page numbering in number
\setcounter{footnote}{0} %
\large
%\linenumbers %%%CD
\section{Introduction}

\begin{figure}[htbp]
\begin{center}
\includegraphics[width=9.0cm]{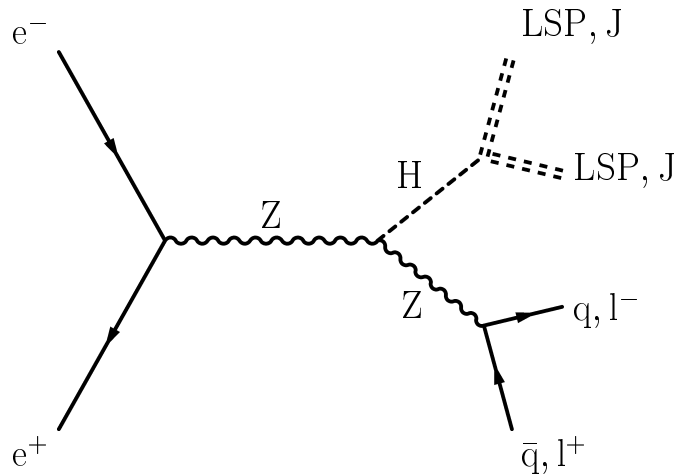}
\caption[]{
Feynman diagram describing the HZ production with the Higgs boson 
decaying into
invisible particles, e.g. the lightest supersymmetric particle (LSP) 
or a Majoron (J) in models with an extended Higgs sector.}
\label{pic:hZinv}
\end{center}
\end{figure}

The data collected by DELPHI have been searched for the presence
of a Higgs boson produced in association with a Z, in $\ee\,\rightarrow\,{\rm HZ}$,
but which decays to stable non-interacting particles.
%A search for the production of $\ee\,\rightarrow\,{\rm HZ}$ was performed. The 
%Higgs was assumed to decay into stable non-interacting particles that do not
%interact in the detector, rendering the Higgs decay invisible.
The process is illustrated in Fig.~\ref{pic:hZinv}.
Such invisible Higgs boson decays can occur in Supersymmetry, 
where the Higgs could decay into a pair of neutralinos
$\tilde{\chi}^0_1$~\cite{Djouadi:1996mj,bij97,khlopov02}.
In such models $\tilde{\chi}^0_1$ is assumed to be the
lightest supersymmetric particle and therefore assumed to be stable. It is weakly interacting
with ordinary matter.
Invisible Higgs decays also occur in Majoron models~\cite{peccei,campos,wells} 
with the Higgs decaying into two Majorons. The results of the search described in this 
article are valid more generally in models with stable Higgs bosons that  do not interact 
in the detector.

Similar searches have been previously performed by 
DELPHI~\cite{DELPHIh96,DELPHIi97} using data at lower centre-of-mass energies and by
other LEP experiments~\cite{Heister:2001kr,Acciarri:2000ec}.
In this paper searches are presented in four different final states, where the Z decays either 
into a $\mathrm{q\bar{q}}$, $\mathrm{e^{+}e^{-}}$, $\mu^{+}\mu^{-}$ or $\tau^{+}\tau^{-}$ pair.

The paper is organised as follows: 
First the analyses in the hadronic channel are addressed separately
in high and low mass ranges. Then we describe the analyses in the 
leptonic channels which cover $\mu$, e, and $\tau$ final states.
Next, the results are summarised and 95\% Confidence Level (CL)
limits are calculated. The limits are then reinterpreted in the 
framework of the Minimal Supersymmetric extension of the Standard
Model (MSSM) and in a Majoron model.

\section{The DELPHI detector and the data set}
The analyses were mainly based on the information from the tracking system, the calorimeters,
the muon chambers, and the photon veto counters of the DELPHI detector. 
%In order to ensure the hermeticity of the detector, 
The scintillation counters veto photons in blind regions of the electromagnetic 
calorimeters at polar angles near $40^\circ$, $90^\circ$ and $140^\circ$.
The DELPHI detector and its performance are described in detail in Ref.~\cite{perfo,Chochula:1997wg}.

The data set analysed in this paper was taken in the years
1998 to 2000. In 1998 and 1999, data were recorded at centre-of-mass
energies 188.7, 191.6, 195.6, 199.6 and 201.7 GeV. In 2000 the LEP
energy was varied from 199.7 to 208.4 GeV and the data
taken at energies below and above 205.8~GeV were analysed as
two independent subsamples, with mean energies of 205.0 and 206.7 GeV.
At the end of the year 2000 data taking, one of the twelve sectors
of the Time Projection Chamber (TPC) became non-operational. Data taken
afterwards were then treated as a separate sample, with a mean
centre-of-mass energy of 206.3 GeV. In the following, these three
subsamples of the 2000 data set will be referred to by the energy
of each simulation for the corresponding data, namely 205.0, 206.5
and 206.5U. The simulation of the last data taking period
(206.5U) included the effect of the missing TPC sector in the detector
setup and the changes in the reconstruction software to partly recover
this loss.

For the analysis of the hadronic and leptonic channels different criteria
are required
on the detector status during data taking. As a result the total data sets  
correspond to 589~\pbinv and 571~\pbinv, respectively.
For the simulation of the signal the {\tt HZHA} generator \cite{HZHA} was 
used for the four final states. For all the years of data-taking simulated 
signal samples with 5000 events per mass point and channel were generated 
with the Higgs masses from
40 to 90~\GeVcc\ in 5 \GeVcc\ steps, from 90 to 115.0~\GeVcc\ 
in 2.5~\GeVcc\ steps and at 120~\GeVcc\ .
%\end{itemize}

The background processes \ee$\rightarrow$\ffbar($n\gamma$) and 
$\rm \mu^+\mu^- (n\gamma)$ were generated using the {\tt KK2F} generator
\cite{Jadach:2000vf} and the background process $\rm \tau^+\tau^- (n\gamma)$ was
generated using the {\tt KORALZ} generator~\cite{Jadach:1997nk}.
The processes which lead to charged and neutral current four-fermion 
final states were generated with the {\tt WPHACT} 
generator~\cite{Accomando:1996es}. The {\tt PYTHIA} generator~\cite{JETSET} 
was used to describe the hadronic two-photon processes and the {\tt BDK}
generator~\cite{bdk} was used to describe the leptonic two-photon processes.
Finally, the {\tt BHWIDE} generator~\cite{Jadach:1994} was used for the Bhabha processes. 
Both signal and background events were processed through the full DELPHI 
detector simulation \cite{perfo}. 
The inoperative sector in the TPC is also taken into account in the corresponding simulation
in the 206.5U data set.

\section{The hadronic channel}
The hadronic decay of the Z represents 70\% of the HZ final states. 
The signature of an invisible Higgs boson decay is a pair of acoplanar 
and acollinear jets with a di-jet mass compatible with the Z mass and missing 
energy and momentum due to the invisibly decaying Higgs boson.

In order to obtain a good performance in the whole mass range, 
two overlapping mass windows were defined for each year of operation and the 
analyses were optimised for each window as defined in Table~\ref{tab:highlowrange}.
\begin{table}
\begin{center}
\begin{tabular}{|c|c|c|}\hline
Year    &Low mass range(\GeVcc) &High mass range(\GeVcc) \\ \hline
1998    &40-90      & 75-120  \\
1999    &40-100      & 75-120  \\
2000    &40-105      & 95-120  \\ \hline
\end{tabular}
\end{center}
\caption{ \label{tab:highlowrange} 
Hadronic channel: Low and high Higgs boson mass ranges for three years of data-taking.}
\end{table}

\subsection{High mass analysis}
The selection of HZ candidate events consists of several steps
in order to suppress the bulk of the background.
First, the events were clustered into jets using the DURHAM~\cite{durham} algorithm. 
Then a preselection was applied to remove most of the two-photon background 
and a great part of the backgrounds due to four-fermion processes and to 
hadronic events with a radiative return to an on-shell Z. Then the final 
separation between the signal and the background channels was achieved 
through an Iterative Discriminant Analysis (IDA)~\cite{IDANIMCPC}. 
The details of the preselection are:

\begin{itemize}
\item Anti-$\gamma\gamma$:
Each event was required to have at least 9 charged particle tracks.
Two of them must have transverse momentum greater than 2 \GeVc\ and 
impact parameters to the primary vertex less than 1 mm in the transverse plane and less 
than 3 mm along the beam axis. 
It was also required that the charged energy be greater than $0.16\sqrt{s}$.
There should be no electromagnetic shower with more than $0.45\sqrt{s}$,
the transverse energy\footnote{The transverse energy is the energy perpendicular to the beam axis, defined as $E_{\mathrm{T}}=\sqrt{p_x^2+p_y^2+m^2}$.}
 be greater than $0.15\sqrt{s}$ and the sum 
of the longitudinal momenta be greater than $0.25 \sqrt{s}$.

\item Anti-\ffbar($n\gamma$) and anti-WW:
A cut in the $\theta_{\mathrm{p_{\rm mis}}}$ vs.
$\sqrt{s^\prime}$ \cite{Abreu:1996va} plane was applied, required
\begin{center}
\begin{description}
%\item   $\rm \theta_{\mathrm{p_{\rm mis}}}  \ge  40^\circ$ and  
%        $\sqrt{s^\prime} \ge 115~\mathrm{GeV}$,
%\item   $\rm \theta_{\mathrm{p_{\rm mis}}} \le 140^\circ$ and 
%        $\sqrt{s^\prime} \ge 115~\mathrm{GeV}$,
\item $40^\circ \le \rm \theta_{\mathrm{p_{\rm mis}}} \le 140^\circ$ and
\item $\sqrt{s^\prime} \ge 115~\mathrm{GeV}$

\end{description}
\end{center}
where $\sqrt{s^\prime}$ stands for the effective
centre-of-mass energy after the initial state radiation of one or more 
photons and 
$\theta_{\mathrm{p_{\rm mis}}}$ is the polar angle of the missing momentum.
In addition, it was required that less than $0.08\sqrt{s}$ was 
deposited in the STIC\footnote{Small angle TIle Calorimeter,
covering the very forward region.} \cite{perfo},
$\sqrt{s^\prime}/ \mathrm{\sqrt{s}}$ 
was less than 0.96 and 
that the total electromagnetic energy within 30$^{\circ}$ of the beam directions was less than 
$0.16\sqrt{s}$.
In order to suppress badly reconstructed events, candidates in which
a jet pointed to the insensitive region between barrel and endcap detectors or 
where both jet axes were below 12$^\circ$ were rejected.
A hermeticity veto algorithm \cite{DELPHIh2000} using the scintillator counters 
was applied to ensure that no photon escaped in the insensitive region of the 
electromagnetic calorimeter at polar angles near $40^\circ$, $90^\circ$ and $140^\circ$.
To suppress background from WW pair production, the energy of the most energetic particle
was required to be less than $0.2\sqrt{s}$ and the transverse momentum of any particle in the jet
with respect to its jet axis (forcing the event into a two-jet configuration) 
to be less than $0.05\sqrt{s}/c$.
Finally, upon forcing the event into a three-jet 
configuration, it was required that every jet had at least one charged particle in order to suppress \ffbar($n\gamma$) events.
\end{itemize}

\label{tailcuts}
Twelve variables were used to construct an effective tagging variable
in the framework of the IDA.
In order to calculate these variables,
the event was forced into two jets.
%, using the DURHAM algorithm.
The variables are:
\begin{itemize}
\item {\EGSSS:} 
the normalised energy of a photon
assumed to have escaped in the beam direction, deduced
from the polar angles of the two jet directions in the event.
The photon energy was normalised to the energy expected
for a photon recoiling against an on-shell~Z.
\item $\ln(p_{\rm T}\;\;\mathrm{[\GeVc]})$:
logarithm of the transverse momentum of the event.
\item $E_{\mathrm{vis}}/\sqrt{s}$:
visible energy of the event, normalised to the centre-of-mass energy.
\item $E_{\mathrm{T}}/\sqrt{s}$:
transverse energy of the event, normalised to the centre-of-mass energy.
\item $\theta_{\rm cone}$: The minimum polar angle defining a cone in the positive and negative beam directions
containing 6\% of the total visible energy.
\item $|\cos\theta_{{p}_{\mathrm{mis}}}|$:
cosine of the polar angle of the missing momentum.
\item $E_{\mathrm{isol}}/\sqrt{s}$:
energy sum between the two cones, defined by half opening angles 5$^\circ$ and
$\alpha_{\rm max}$ around the most isolated particle. The energy sum is then normalised to the centre-of-mass energy.
The most isolated particle is defined as the particle with momentum above 
2~\GeVc\ with the smallest energy sum in the double cone. 
In the momentum interval from 2 to 5~\GeVc, $\alpha_{\rm max}$ is set to 
60$^\circ$ in order to maximise the sensitivity to isolated particles 
from tau decays in
$\mathrm{WW}\rightarrow\mathrm{q\bar{q}}^\prime\tau\nu$ events. 
An opening angle of 25$^\circ$ is used for particles
with momenta above 5~\GeVc. 
\item $p_{\mathrm{isol}}/\sqrt{s}$:
momentum of the most isolated particle, as defined above, normalised to the centre-of-mass energy.
\item $\rm \log_{10} (scaled\;\; acoplanarity)$:
The acoplanarity is defined as $180^\circ - \Delta\phi$, where
$\Delta\phi$ is the difference in azimuthal angle 
(in the plane perpendicular to the beam axis) between the two
jets. 
In order to compensate for the geometrical instability of the acoplanarity 
for jets at low angles it was multiplied with the angle between the two jets.
\item Thrust:
thrust value of the event, computed in the rest frame of the
visible system. 
\item $\rm \ln(acollinearity)$: 
logarithm of the acollinearity (in degrees) of the two-jet system.
\item $\ln(\max(p_{\rm T}\;\;\mathrm{[\GeVc]})_{\rm jet})$: 
highest transverse momentum of the jet-particles,
defined by the transverse momentum of any particle in the jet
with respect to the jet axis.
\end{itemize}

\begin{table}
\begin{center}
\begin{tabular}{|l|c|c|} \hline
Variable                & lower cut     & upper cut \\ \hline
\EGSSS                  & -           & 0.90      \\
${\ln(p_{\mathrm{\rm T}}\;\;\mathrm{[\GeVc]})}$ & 1.75          & 4.5       \\
${E_{\mathrm{\rm T}}/\sqrt{s}}  $   &  0.15 & 0.6       \\
${p_{\rm isol}/\sqrt{s}}$&  0.008  & 0.18      \\
$\rm \log_{10} (scaled\;\;acoplanarity)$      &  0.3       & 2.5      \\
Thrust                     & 0.65       & 1.0       \\
$\rm \ln(acollinearity)$& 2.0      & 4.5       \\
$\ln(\max(p_{\rm T}\;)_{\rm jet}\;\;\mathrm{[\GeVc]})$
                          & $-0.5$      & 2.50      \\ \hline
\end{tabular}
\caption{ \label{tab1} Tail cuts used in the high mass hadronic analysis.
The variables are described in detail in section~\ref{tailcuts}.}
\end{center}
\end{table}

\begin{figure}[htbp]
\begin{center}
\subfigure[]{
\includegraphics[width=7.5cm]{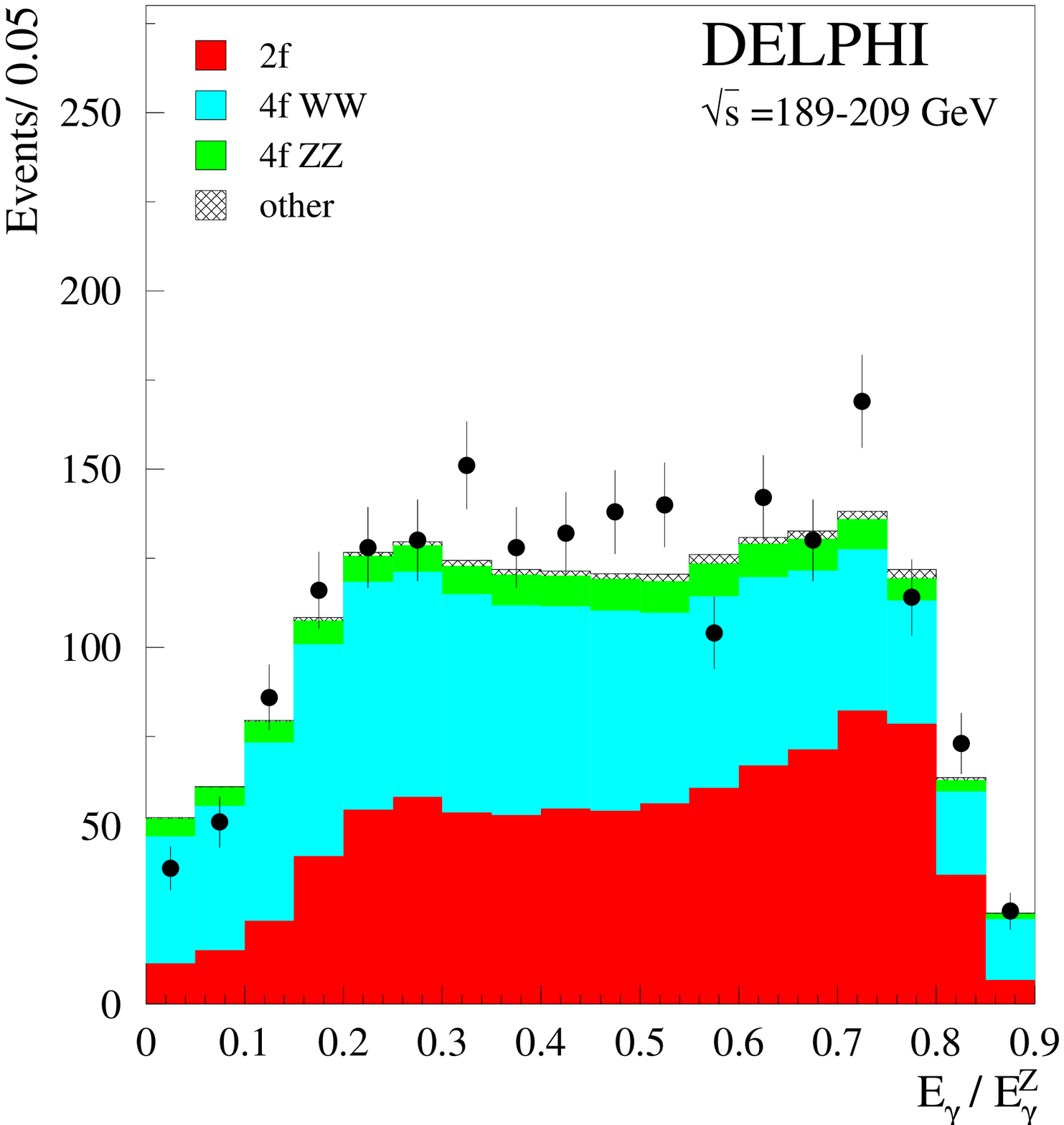}
}
\subfigure[]{
\includegraphics[width=7.5cm]{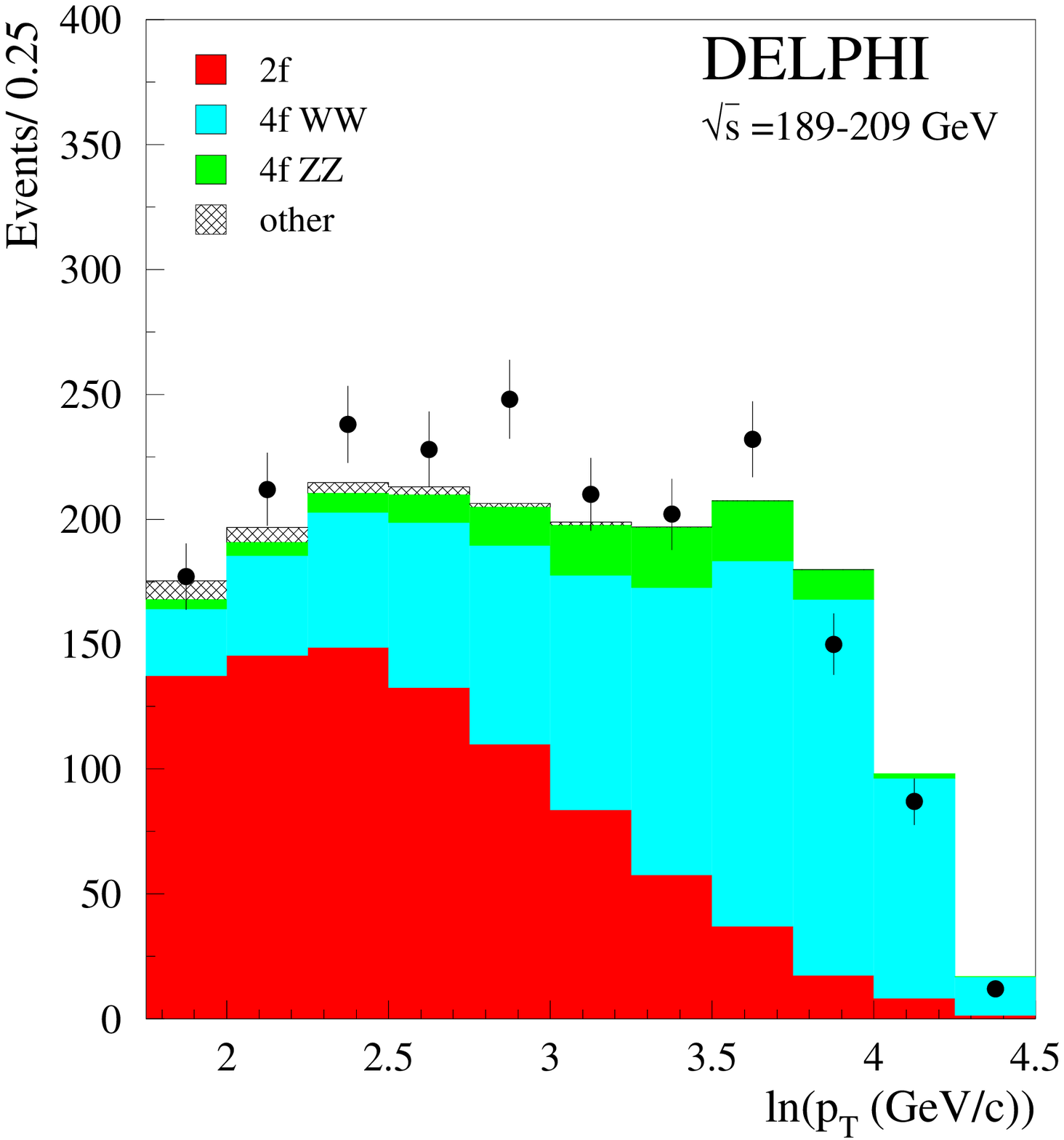}
}
\subfigure[]{
\includegraphics[width=7.5cm]{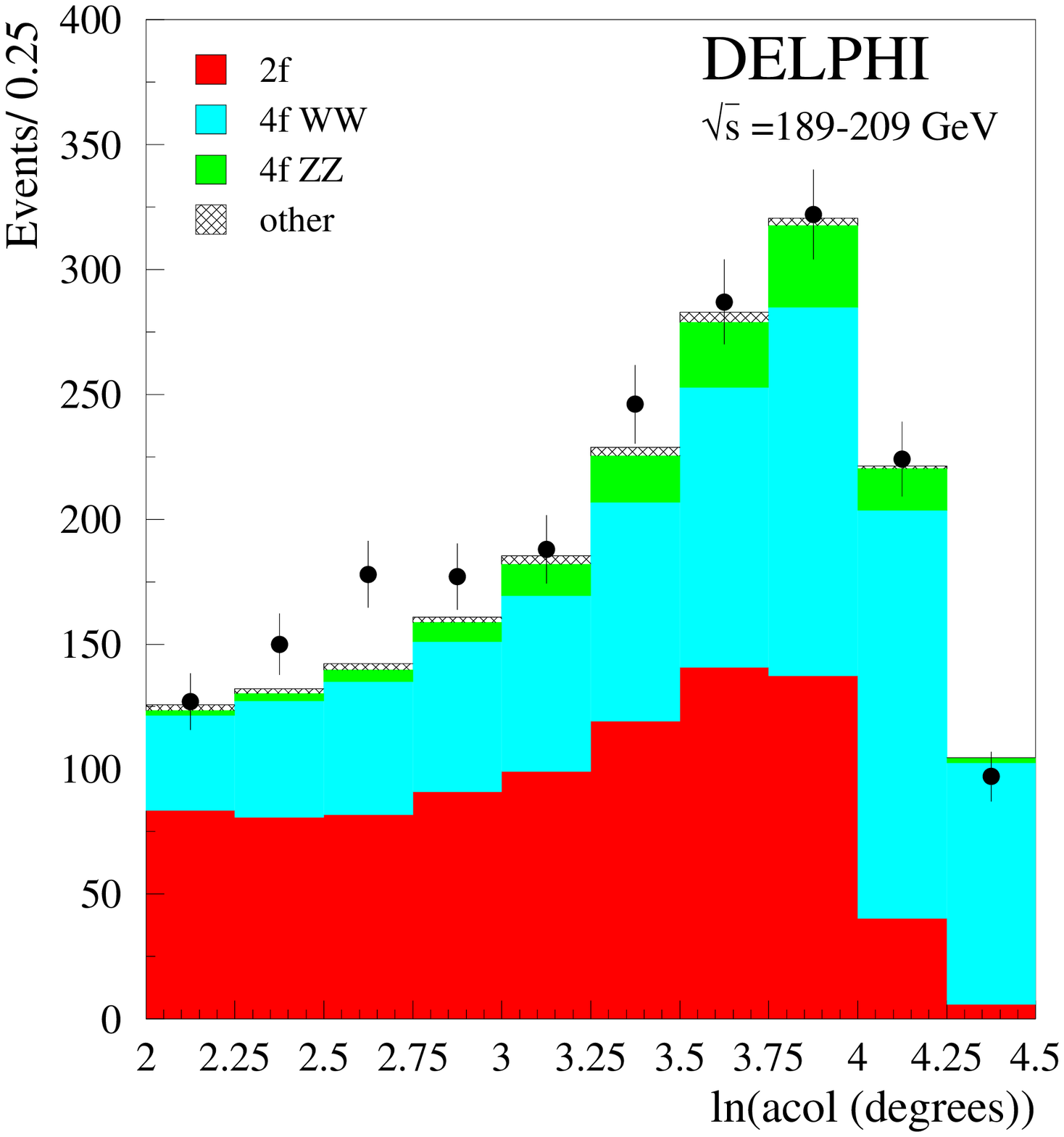}
}
\subfigure[]{
\includegraphics[width=7.5cm]{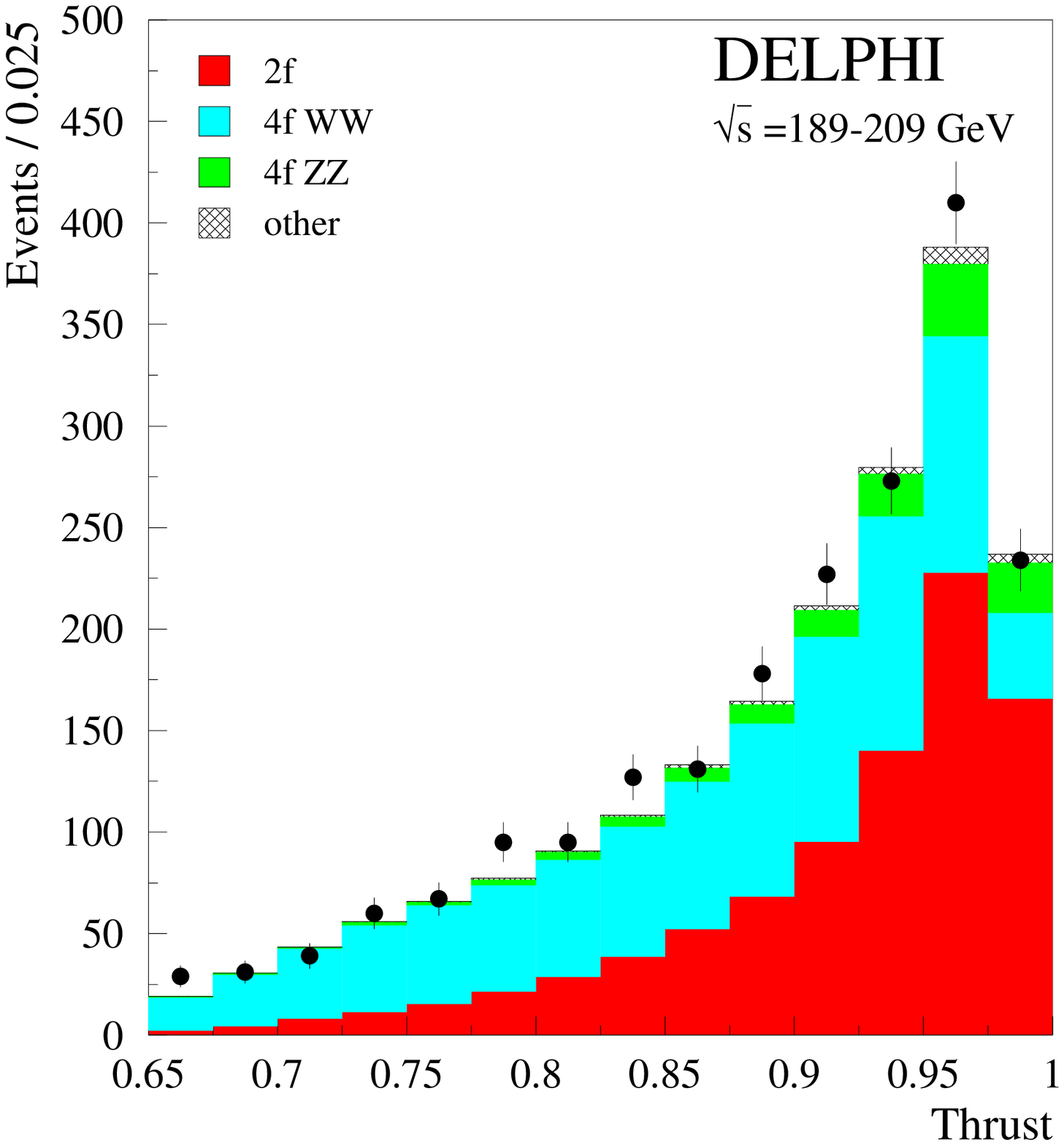}
}
\end{center}
\caption{\label{fig:presel_hinvqq} 
Hadronic channel high mass analysis: 
Distribution of the four IDA input variables after the final
preselection as described in section~\ref{tailcuts}:
%Distributions for the preselection variables 
%after the tail cuts as described in section~\ref{tailcuts}: 
a) \EGSSS ; b) ln$(p_{\rm T})$ in~\GeVc ; c) ln(acollinearity);  d) Thrust.}
\end{figure}

The cuts listed in Table~\ref{tab1} were applied in the tails 
of the distribution of these variables in order to concentrate on the 
signal region and to avoid long tails in the input variables for the
IDA. In addition to the cuts listed in Table~\ref{tab1}, the number of 
electrons or muons identified by the standard DELPHI algorithms \cite{perfo} was required to be less than three.
The agreement between data and simulation is shown in 
Table~\ref{tab:tabpresel} and Fig.~\ref{fig:presel_hinvqq}. 
There is a small excess in data over expected background, which is not concentrated in one bin.
\begin{table}[h]
\begin{center}
\begin{tabular}{|c|c|c|c|c|c|c|}\hline
$\mathrm{\sqrt{s}} $& \multicolumn{2}{|c|}{Anti-$\gamma\gamma$} & \multicolumn{2}{|c|}{Anti-\ffbar($n\gamma$) \& anti-WW} & \multicolumn{2}{|c|}{Tail cuts}  \\
(GeV)&Data  &MC             &Data &MC            &Data&MC            \\\hline
188.6 &15115 &$14967.0\pm8.1$&1578 &$1565.2\pm6.0$&494 &$485.9\pm3.2$ \\\hline
191.6 &2394  &$2351.8\pm1.3$ &258 &$249.9\pm0.9$ & 88 &$ 79.0\pm0.5$ \\\hline
195.5 &7040  &$6782.4\pm3.7$ &739 &$734.9\pm2.8$ &242 &$242.0\pm1.4$ \\\hline
199.5 &7296  &$7168.9\pm3.9$ &784 &$795.4\pm2.8$ &295 &$264.4\pm1.6$ \\\hline
201.6 &3557  &$3407.8\pm1.9$ &396 &$382.9\pm1.3$ &152 &$130.6\pm0.7$ \\\hline
205.0 &6272  &$6011.6\pm3.7$ &678 &$686.2\pm2.4$ &240 &$239.2\pm1.3$ \\\hline
206.5 &6772  &$6697.0\pm4.5$ &798 &$768.5\pm2.9$ &283 &$268.2\pm1.6$ \\\hline
206.5U&4472  &$4560.4\pm3.9$ &534 &$541.5\pm2.6$ &202 &$190.7\pm1.5$ \\\hline
\end{tabular}
\end{center}
\caption{Comparison of simulation and data after the different steps 
of the preselection in the high mass hadronic analysis.
The listed errors are from Monte Carlo statistics only.
\label{tab:tabpresel}}
\end{table}

\begin{figure}[htbp]
\begin{center}
\subfigure[]{
\includegraphics[width=8.0cm]{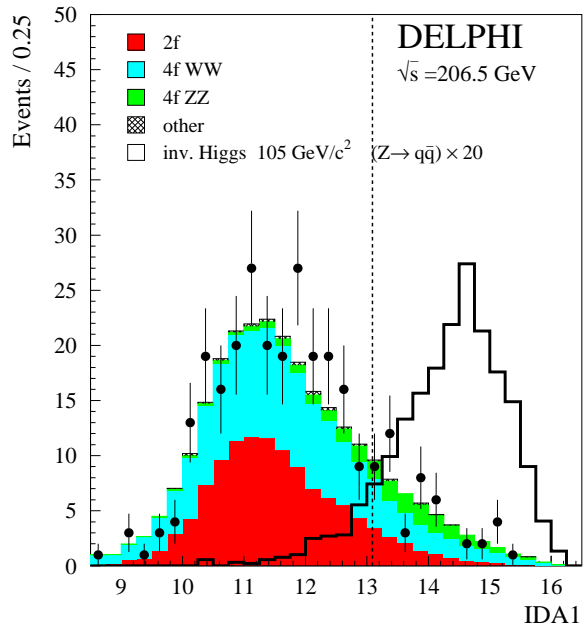}
}
\subfigure[]{
\includegraphics[width=8.0cm]{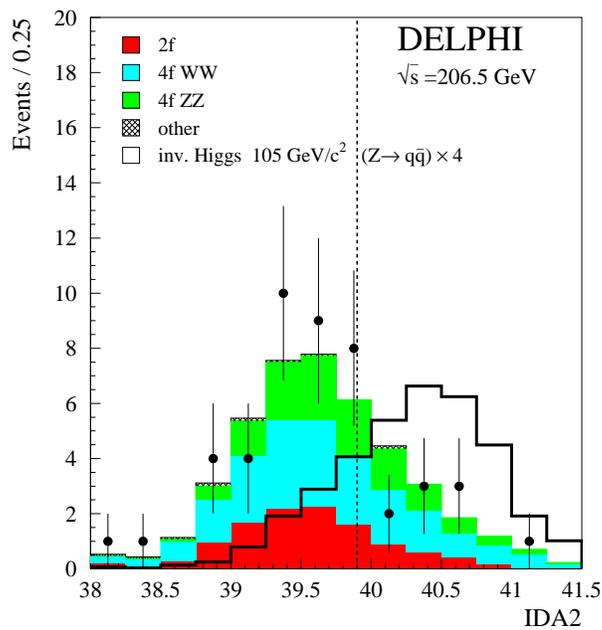}
}
\end{center}
\caption{\label{fig:ida_hinvqq} 
Hadronic channel high mass analysis: Distributions for the IDA variables after first 
(a) and second IDA step (b) at $\sqrt{s}=206.5$~GeV. 
The dashed line indicates the cut on the IDA variable. 
The white histogram shows the expectation of a 105~\GeVcc~Higgs
signal where the signal rate is enhanced by a factor 20 for (a) and 4 for (b).}
%with an arbitrary scale.}
\end{figure}

The IDA is a modified Fisher Discriminant Analysis, the two main differences are the introduction of a non-linear
discriminant function and iterations in order to enhance the separation of signal and background.
Two IDA steps were performed, with a cut after the first IDA iteration keeping 90 \% of the 
signal efficiency.  In order to have two independent samples for the derivation of the IDA function
and for the expected performance, the signal and background samples were divided in two equally sized samples.
As an illustration, the distributions of the two IDA variables at $\sqrt{s}=206.5$ GeV are shown in 
Fig.~\ref{fig:ida_hinvqq}. The slight disagreement in the rates observed at the preselection level is effectively removed by the 
IDA analysis, since it is concentrated mostly outside the signal region.

\begin{figure}[htbp]
\begin{center}
\includegraphics[width=15cm]{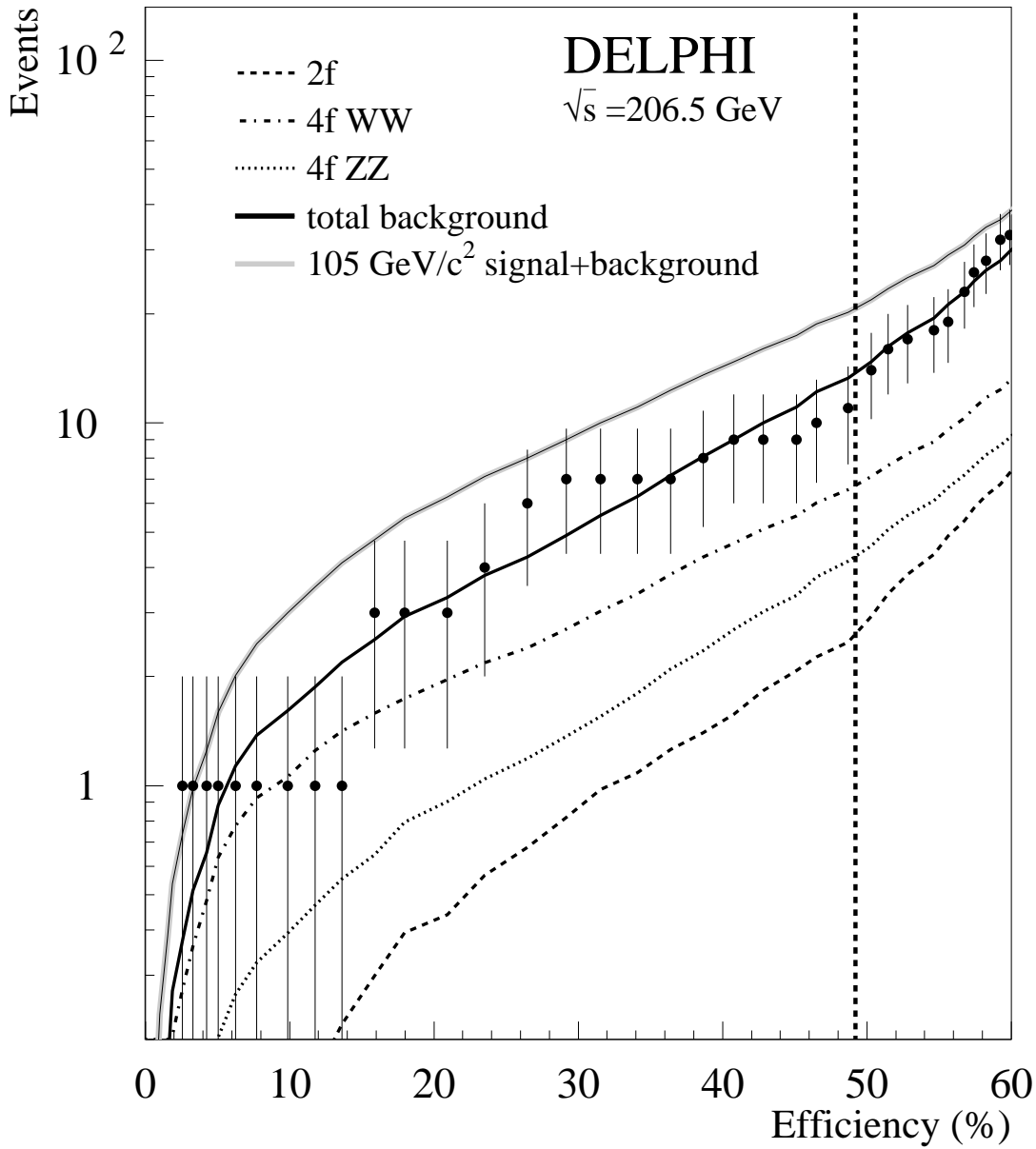}
\caption[]{
Hadronic channel high mass analysis:
Data and expected background for the 206.5 GeV centre-of-mass 
energy as a function of the efficiency for an invisibly decaying Higgs
boson of 105~\GeVcc. The lines represent the most important backgrounds with the solid black line 
showing the sum of all the background processes. In addition the grey line shows 
the expectation for a 105~\GeVcc~Higgs signal added on top of the background. 
The vertical dashed line indicates the final cut chosen to maximise the sensitivity.}
\label{pic:effq}
\end{center}
\end{figure}

The observed and expected rates at $\sqrt{s}=206.5$ GeV are shown in Fig.~\ref{pic:effq} as a function of the efficiency to detect 
a 105 \GeVcc\ Higgs boson when varying the cut on the second IDA variable.
The final cut on the second IDA variable was determined by maximising 
the expected exclusion power. This was done separately for each centre-of-mass energy to optimise 
the analysis for a 85~\GeVcc~Higgs boson at 188.6 GeV,
for a 95~\GeVcc~Higgs boson at 191.6 and 195.6~GeV, 
for a 100~\GeVcc~Higgs at 199.5 and 201.6 GeV and 
for a 105~\GeVcc~Higgs at 205.0, 206.5 GeV and 206.5U GeV. Here we assume SM production cross-section and 
a branching ratio of 100\% into invisible final states.
For example, in Fig.~\ref{pic:effq} the cut on the second IDA is indicated by the dashed vertical line.
The final number of selected events in data and Monte Carlo simulations is given in 
Table~\ref{tab:tabeff}.

\subsection{Low-mass analysis}
\begin{figure}[htbp]
\begin{center}
\subfigure[]{
\includegraphics[width=7.5cm]{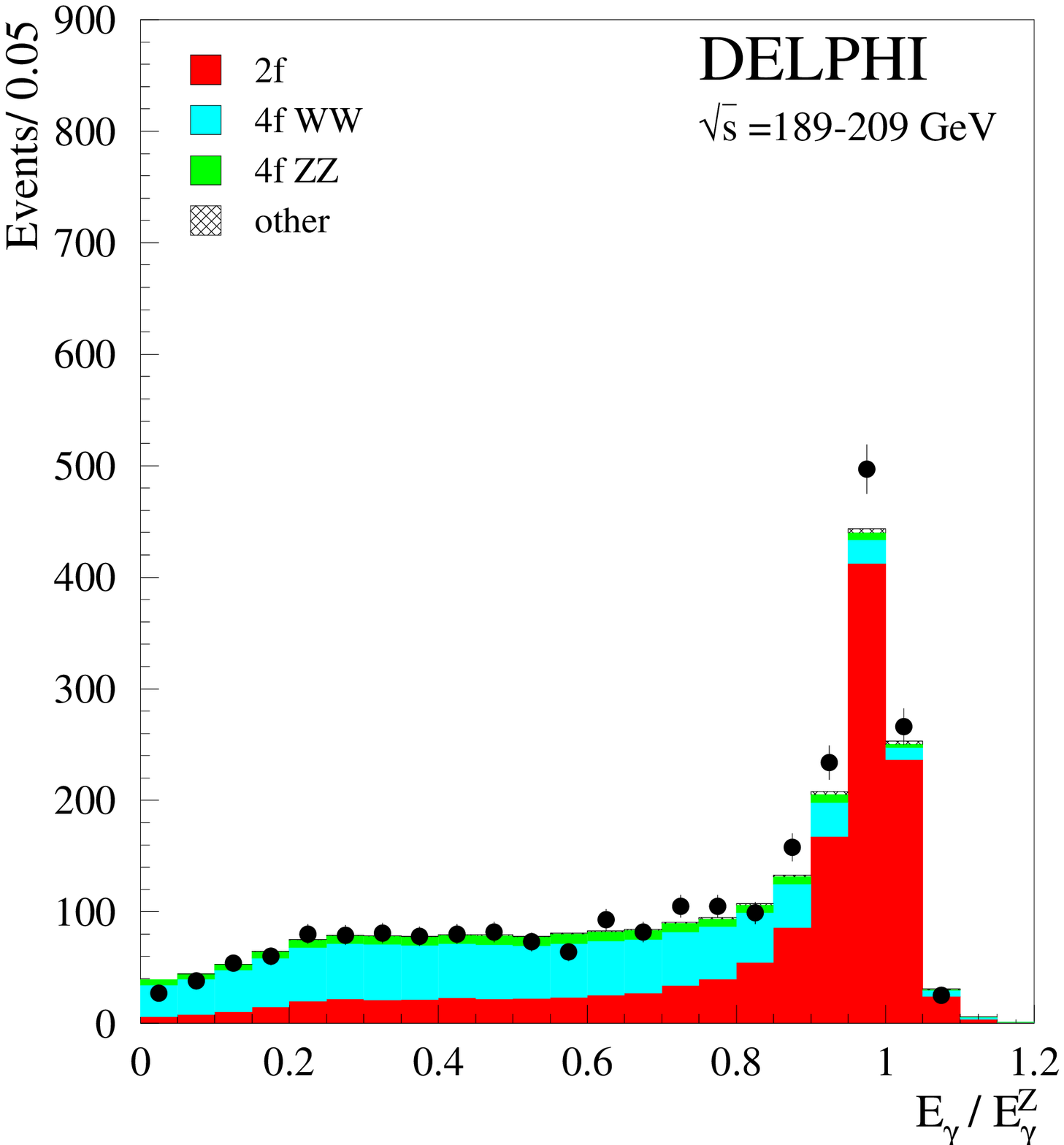}
}
\subfigure[]{
\includegraphics[width=7.5cm]{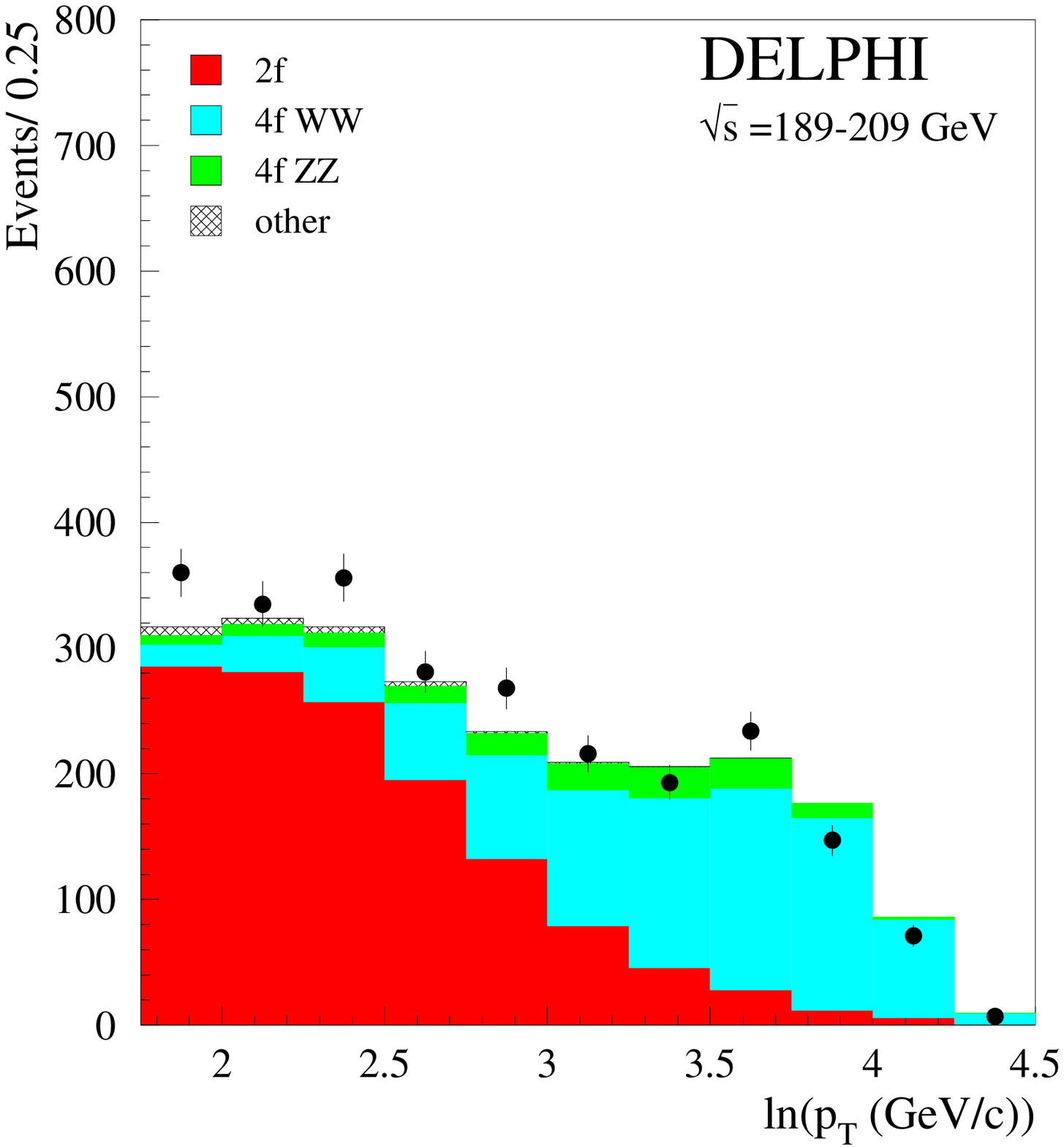}
}
\subfigure[]{
\includegraphics[width=7.5cm]{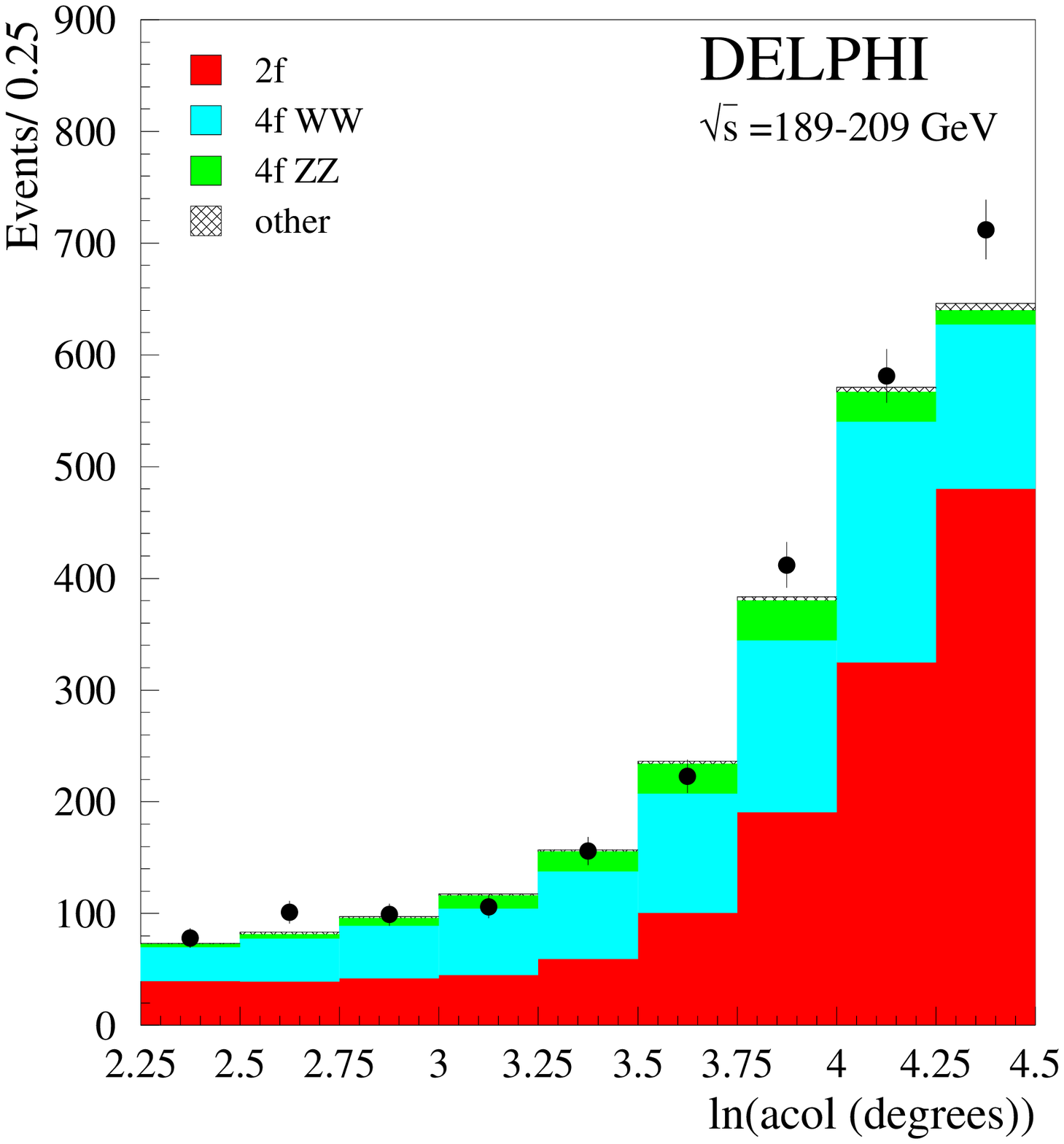}
}
\subfigure[]{
\includegraphics[width=7.5cm]{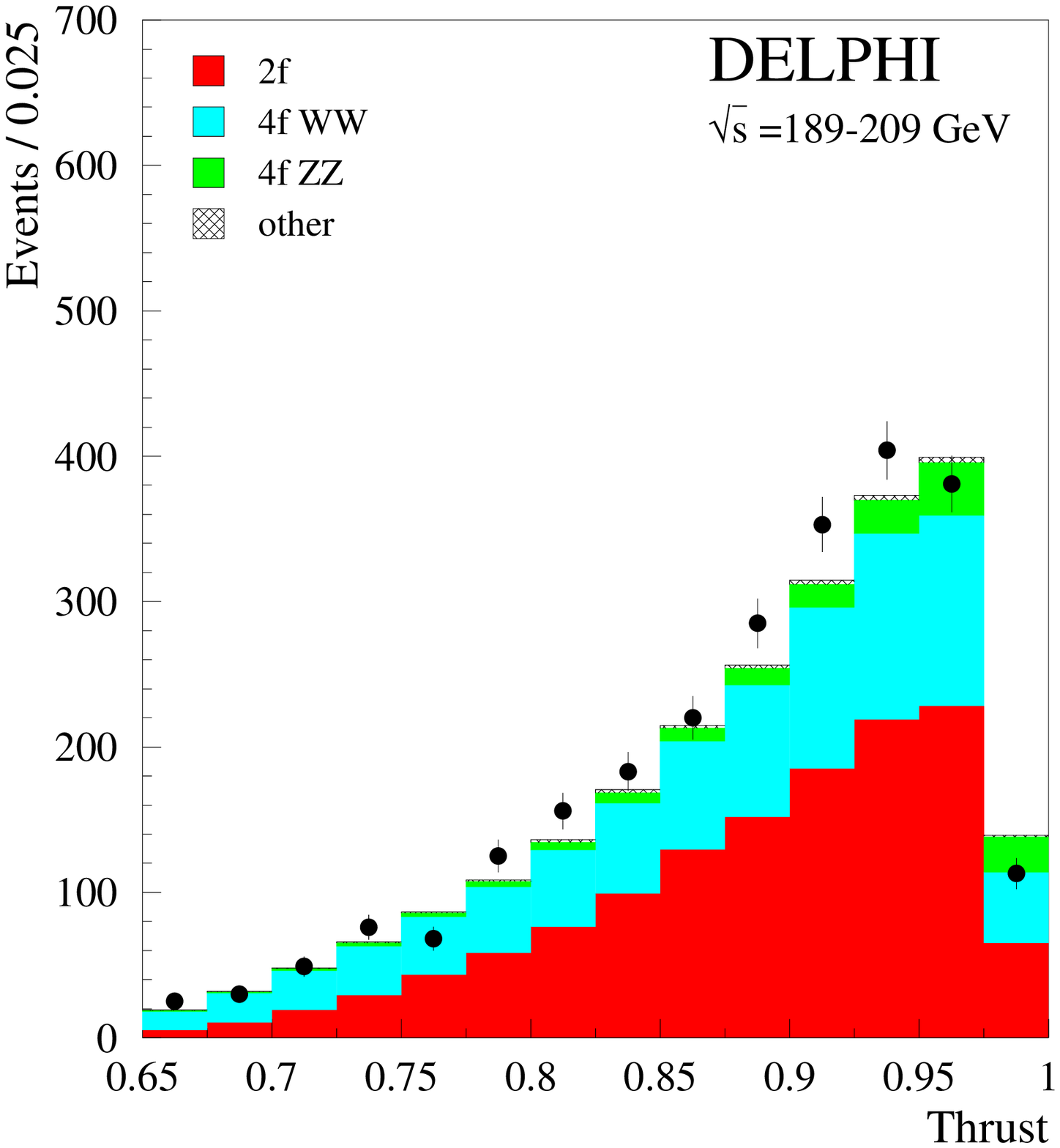}
}
\end{center}
\caption{\label{fig:presel_hinvqq-l}
Hadronic channel low mass analysis: 
Distribution of the four IDA input variables after the final
preselection as described in section~\ref{tailcuts}:
%Distributions for the preselection variables 
%after the tail cuts as described in section~\ref{tailcuts}: 
a) \EGSSS ; b) ln$(p_{\rm T})$ in~\GeVc ; c) ln(acollinearity); d) Thrust.}
\end{figure}

For the low-mass analysis, the preselection was adapted for the different event shape and kinematics.
In the anti-\ffbar($n\gamma$) and anti-WW selection the cut in the 
$\theta_{\mathrm{p_{mis}}}$ vs $\sqrt{s^\prime}$ plane
and the cut on $\sqrt{s^\prime}/ \mathrm{\sqrt{s}}$
were removed in order to increase the signal efficiency.
This was possible because the signal events have a much smaller amount of missing energy than the 
events in the high-mass range.
Some tail cuts were also slightly changed as shown 
in Table~\ref{tab1a} and a cut requiring 
the visible mass to be at least 20\% of $\sqrt{s}$ was added.
Figure~\ref{fig:presel_hinvqq-l} and Table~\ref{tab:tabpresel-low} show the agreement of data and background at the 
preselection level. Figure 5 a) shows an excess
of data over the expected background near \EGSSS =1 due to
an underestimation of the two-fermion processes.
%Z contribution. 
This region is effectively removed
by the IDA analysis.

\begin{table}
\begin{center}
\begin{tabular}{|l|c|c|} \hline
Variable                        & lower cut & upper cut \\ \hline
\EGSSS                          & -         & 1.20      \\
${E_{\mathrm{\rm T}}/\sqrt{s}}$ & -       & 0.6       \\
${p_{\rm isol}/\sqrt{s}}$       & -         & 0.18      \\
$\rm \log_{10} (scaled\;\;acoplanarity)$    &1.0       & 2.5       \\
$\rm \ln(acollinearity)$        & 2.25      & 4.5       \\ \hline
\end{tabular}
\caption{ \label{tab1a} Tail cuts used in the low mass hadronic analysis.
The variables are described in detail in section~\ref{tailcuts}.}
\end{center}
\end{table}

\begin{figure}[htbp]
\begin{center}
\subfigure[]{
\includegraphics[width=8.0cm]{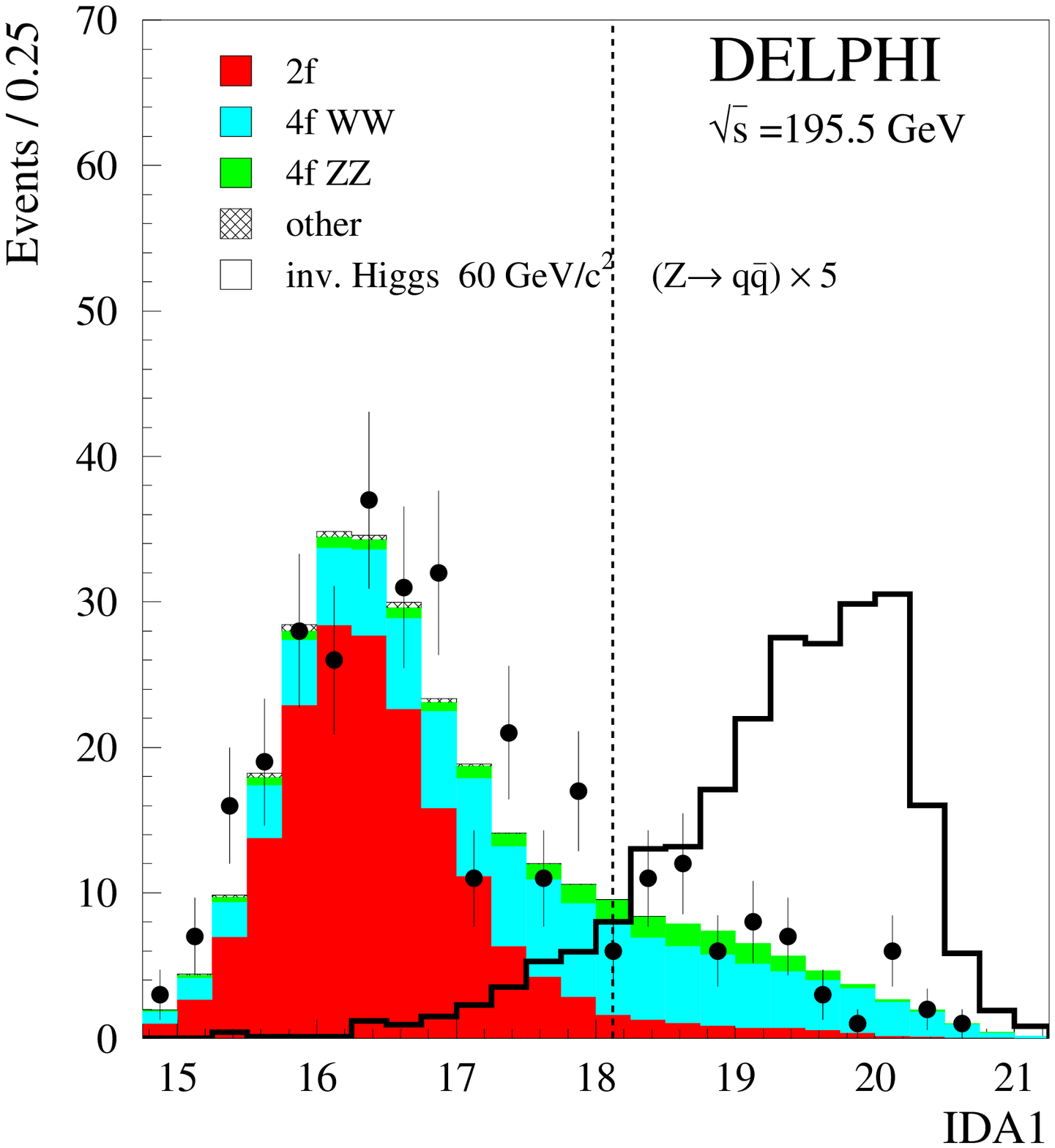}
}
\subfigure[]{
\includegraphics[width=8.0cm]{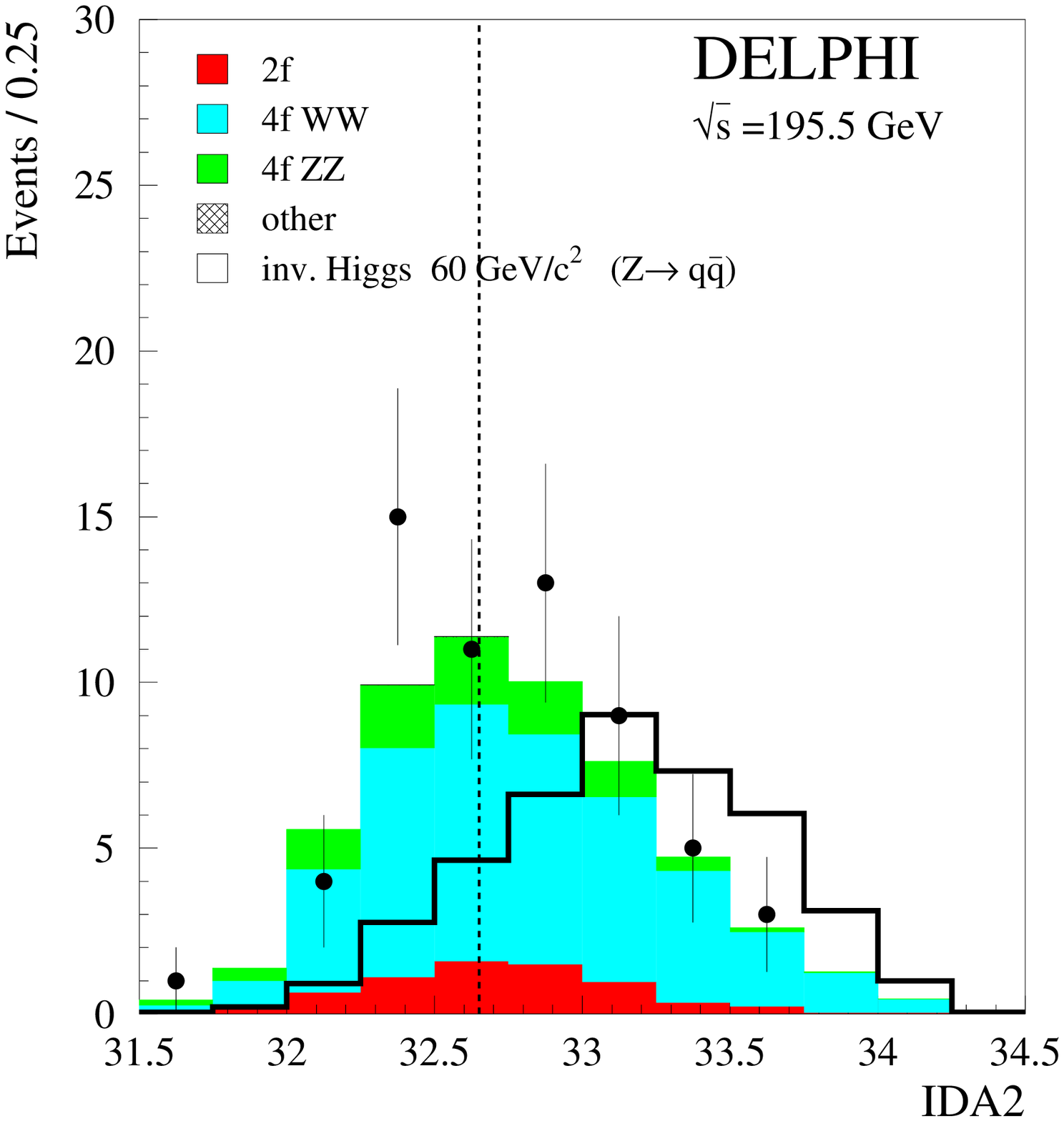}
}
\end{center}
\caption{\label{fig:ida_hinvqq-l}
Hadronic channel low mass analysis: Distributions for the IDA variables after first
(a) and second IDA step (b) at $\sqrt{s}$=195.5 GeV.
The dashed line indicates the cut on the IDA variable.
The white histogram shows the expectation of a 60~\GeVcc~Higgs
signal where the signal rate is enhanced by a factor 5 for (a).}
\end{figure}

\begin{figure}[htbp]
\begin{center}
\includegraphics[width=15cm]{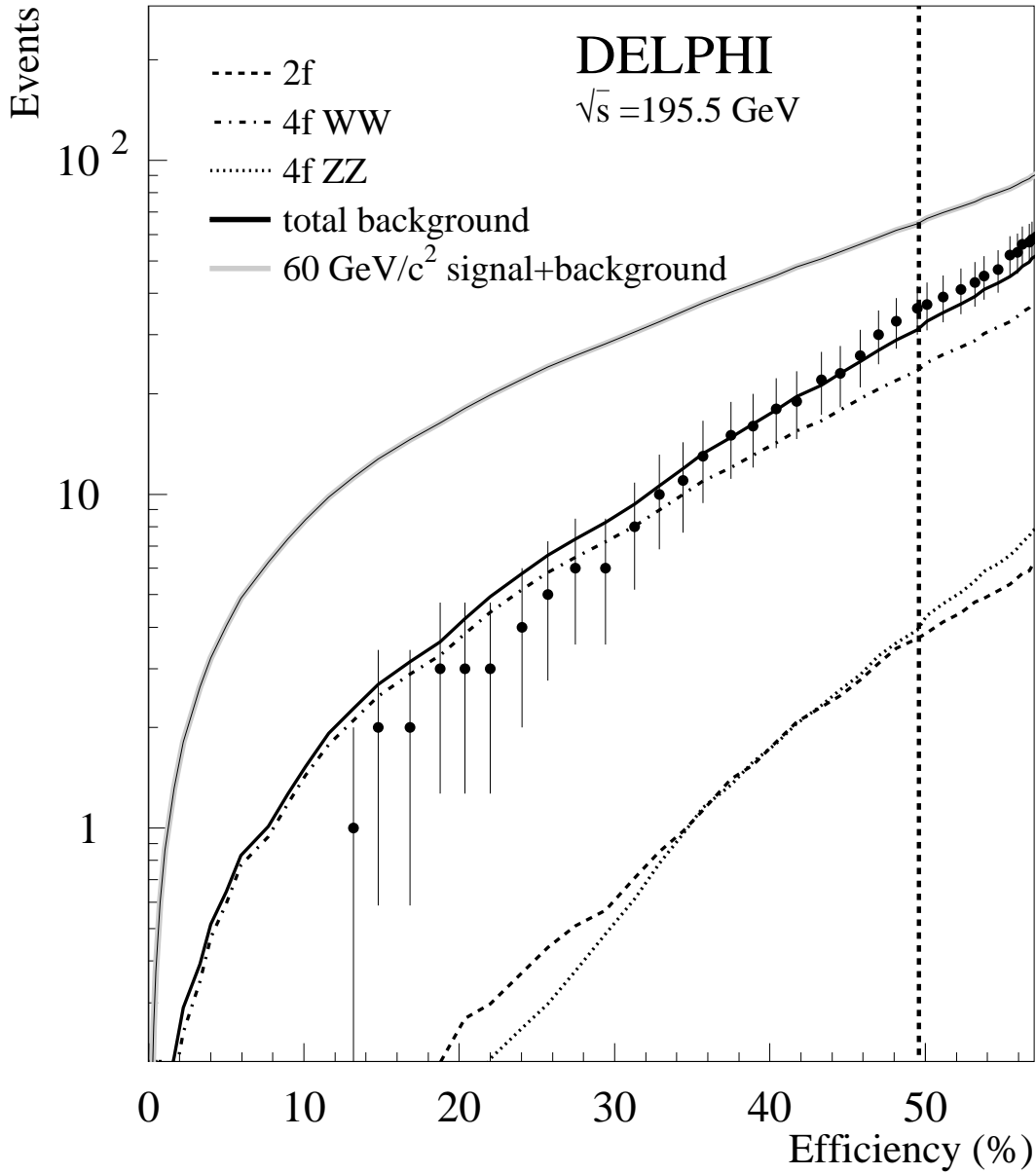}
\caption[]{
Hadronic channel low mass analysis:
Data and expected background for the 195.5 GeV centre-of-mass 
energy as a function of the efficiency 
for an invisibly decaying Higgs boson of 60~\GeVcc. 
The lines show number of events from the most important background 
reactions and the solid black line shows the sum of all the background 
processes. 
In addition the grey line shows the expectation for a 60~\GeVcc~Higgs 
signal added on top of the background. 
The vertical dashed line indicates the final cut chosen to maximise 
the sensitivity.}
\label{pic:effq-l}
\end{center}
\end{figure}

\begin{table}[h]
\begin{center}
\begin{tabular}{|c|c|c|c|c|c|c|}\hline
$\mathrm{\sqrt{s}}$     & \multicolumn{2}{|c|}{Anti-$\gamma\gamma$} & \multicolumn{2}{|c|}{Anti-\ffbar($n\gamma$) \& anti-WW} & \multicolumn{2}{|c|}{Tail cuts}  \\
(GeV)&Data   &MC             &Data   &MC               &Data &MC                       \\ \hline
188.6&15115&$14967.0\pm8.1$&6604 &$6735.2\pm11.0$&622 &$652.0\pm3.9$ \\\hline
191.6& 2394&$ 2351.8\pm1.3$&1013 &$1051.2\pm 1.7$&112 &$103.0\pm0.6$ \\\hline
195.5& 7040&$ 6782.4\pm3.7$&2939 &$3003.0\pm 4.8$&322 &$301.3\pm1.8$ \\\hline
199.5& 7296&$ 7168.9\pm3.9$&3122 &$3117.7\pm 5.0$&338 &$315.1\pm1.8$ \\\hline
201.6& 3557&$ 3407.8\pm1.9$&1551 &$1495.9\pm 2.4$&168 &$152.1\pm0.8$ \\\hline
205.0& 6272&$ 6011.6\pm3.7$&2617 &$2614.9\pm 4.3$&344 &$307.3\pm1.6$ \\\hline
206.5& 6772&$ 6697.0\pm4.5$&2885 &$2909.0\pm 5.2$&305 &$293.6\pm1.7$ \\\hline
206.5U&4472&$ 4560.4\pm3.9$&1878 &$1982.5\pm 4.7$&257 &$237.5\pm1.7$ \\\hline
\end{tabular}
\end{center}
\caption{Comparison of simulation and data after the different steps 
of the preselection in the low mass hadronic analysis.
The errors given are from Monte Carlo statistics only.
\label{tab:tabpresel-low}}
\end{table}
The low-mass analysis also used two IDA steps in order to obtain optimal signal to background
discrimination. 
The distributions of the two IDA variables at $\sqrt{s}$=195.5 GeV are shown in 
Fig. \ref{fig:ida_hinvqq-l}. The observed and expected rates at $\sqrt{s}=195.5$ GeV are shown in 
Fig.~\ref{pic:effq-l} as a function of the efficiency to detect a Higgs boson when 
varying the cut on the second IDA variable.
The cut on the second IDA variable was again determined separately for each centre-of-mass energy as described above.
It was optimised for a 60~\GeVcc~Higgs boson mass at all energies.
The final number of selected events in data and Monte Carlo simulations is given in Table~\ref{tab:tabeff-l}.

\subsection{Mass reconstruction}
\begin{figure}[htbp]
\begin{center}
\includegraphics[width=14.0cm]{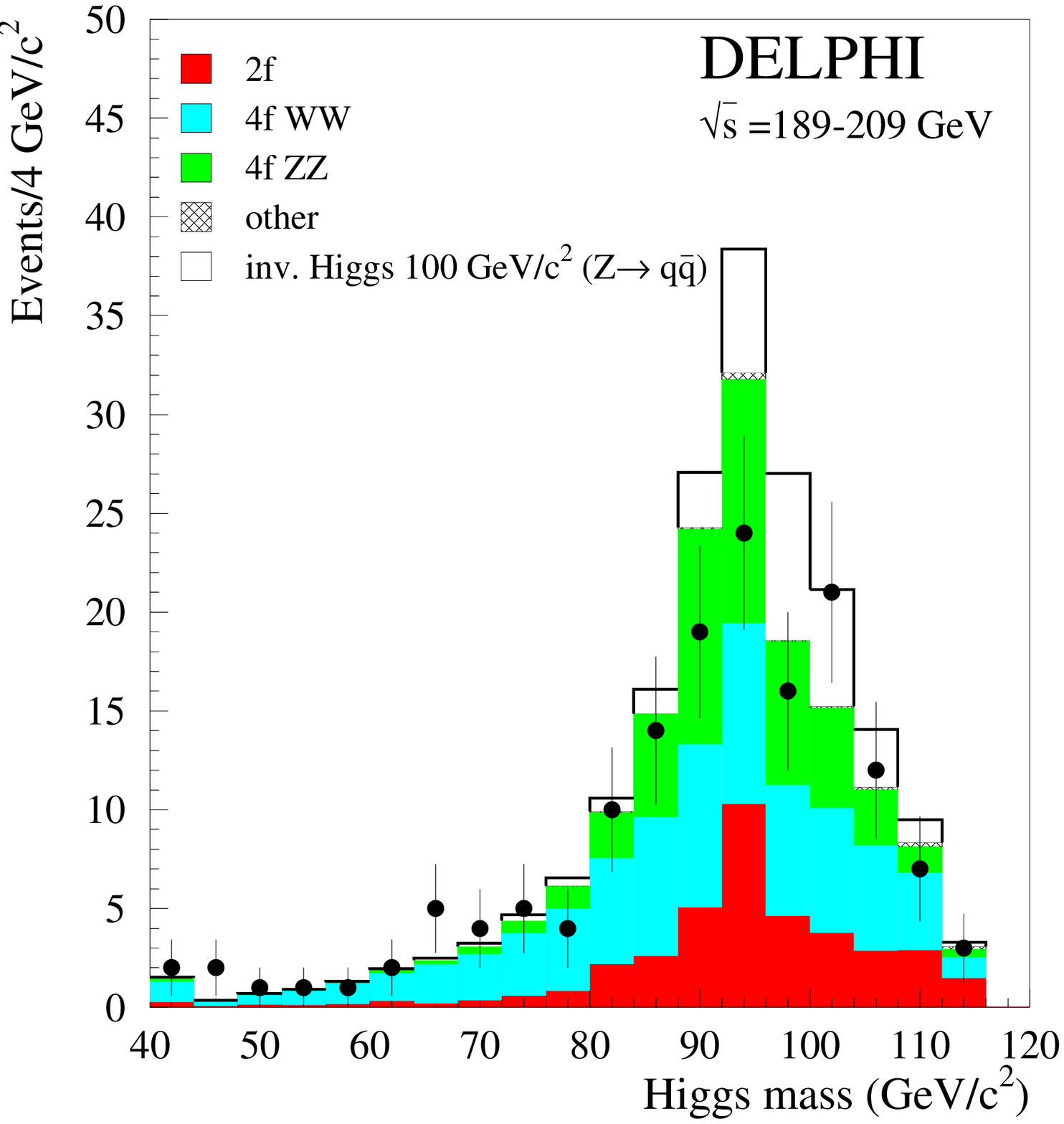}
\end{center}
\caption{\label{fig:massqq} 
Hadronic channel high mass analysis: Reconstructed Higgs boson mass 
for $\sqrt{s}$ from 189 to 209~\GeV\ after the final selection. 
The white histogram corresponds to a Higgs boson with 100~\GeVcc\ mass decaying 
with a branching fraction of 100\% into invisible modes.}
\end{figure}

\begin{figure}[htbp]
\begin{center}
\includegraphics[width=14.0cm]{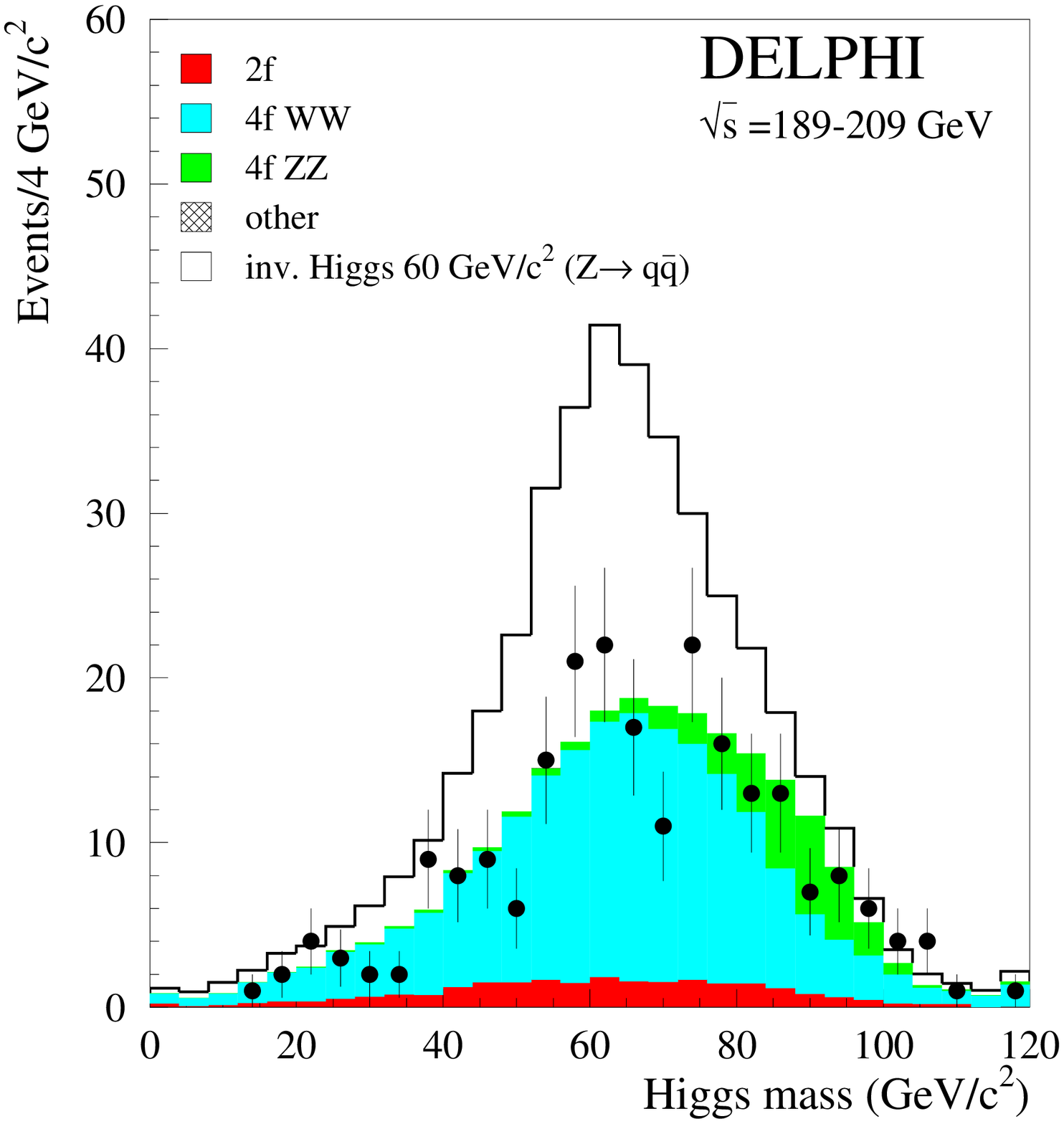}
\end{center}
\caption{\label{fig:massqq-l} 
Hadronic channel low mass analysis: Reconstructed Higgs boson mass
for $\sqrt{s}$ from 189 to 209~\GeV\ after the final selection. 
The white histogram corresponds to a Higgs boson with 60~\GeVcc\ mass decaying 
with a branching fraction of 100\% into invisible modes.}
\end{figure}

The recoil mass to the di-jet system corresponds to the mass of the invisible Higgs boson.
It was calculated with a Z mass constraint for the measured di-jet system
from the visible energy $\Evis$ and the visible mass $\Mvis$. The following expression was used
\begin{displaymath}
  {m_{\mathrm {inv}}} =
  \sqrt{ \left(\mathrm{\sqrt{\it s}} - \frac{\MZ\Evis}{\Mvis}\right)^2
    -\left(\frac{\MZ {p_{\mathrm{mis}}}}{\Mvis}\right)^2 },
\end{displaymath}
where $p_{\mathrm{mis}}$ is the missing momentum and \MZ~is the Z mass. 
The recoil mass distribution after the final selection for the high-mass 
analysis is shown in Fig.~\ref{fig:massqq}.
For the low-mass region this method was also used. In cases where the fit obtained
negative mass squares the standard missing mass calculation 
$\sqrt{E^2_{\rm mis}-p^2_{\rm mis}}$ was used, where 
$E_{\rm mis} = \sqrt s - E_{\rm vis}$.
The recoil mass distribution for the low mass analysis is shown in Fig.~\ref{fig:massqq-l}.

\subsection{Systematic errors}
Several sources of systematics have been considered, first
the effect of modelling the \ffbar($n\gamma$) background 
from different generators was studied by replacing the 
{\tt KK2F} generator with the {\tt ARIADNE} generator~\cite{ariadne} 
at 206.5 GeV.
The results were identical within statistical errors. 
The error of the luminosity is conservatively taken to be $\pm$0.5\%.
The process $\ee\rightarrow\qqev$ provides about a fifth of the background
and the uncertainty on the cross-section of this process is taken to 
be $\pm$5\%~\cite{Grunewald:2000ju}.
This leads to an $\pm$1\% uncertainty of the background.
In order to see the influence of the jet clustering algorithm the DURHAM 
algorithm was replaced by the LUCLUS algorithm \cite{LUCLUS}. 
This results in an uncertainty on the background estimation and 
the signal efficiency in the order of $\pm$1\% for the 
high mass regime and an error in the order of $\pm$2.5\% for the low mass regime.

The data and Monte Carlo simulation were found to be in good overall
agreement. However, since we are searching for events with a large
amount of missing energy, we become sensitive to the tails of the
distributions from the expected Standard Model background events.
When analysing the same topology for the measurement of the Z pair 
production cross-section~\cite{ZZ}, it was found that the small disagreement in 
the tails can be cured if the particle multiplicities of data and Monte Carlo 
simulation are brought into agreement. 
In order not to bias the present analysis, where the disagreements in the 
tails could come from new physics, the tuning of the particle multiplicities 
was done with $\mathrm{Z\rightarrow q\bar{q}}$ events taken at 
$\sqrt{s}$ = 91.1 GeV for each year of data taking.  
The particle multiplicities were estimated separately for the barrel
($\cos\theta\le0.7$) and the forward region ($\cos\theta > 0.7$) and for
different momentum bins and separately for neutral and charged particles. For each of these classes of
multiplicities a separate correction factor $P$ was calculated using
\begin{displaymath}
P=\frac{<N_{\rm data}>-<N_{\rm MC}>}{<N_{\rm MC}>},
\end{displaymath}
where $<N_{\rm data}>$ is the mean value of the particle multiplicity in the data and
$<N_{\rm MC}>$ is the corresponding simulated value.
These correction factors are of the order of a few percent in the barrel
region, they tend to be larger in the forward region and are also larger
for neutral than for charged particles. The correction factors obtained
were then applied to the high energy LEP2 Monte Carlo
simulation on an event by event basis.
The factor $P$ was used as a probability to modify the particle multiplicities 
in the Monte Carlo simulation and related variables were recalculated.
If $P$ was less than zero, there were fewer particles in data than in Monte
Carlo and the particles of the corresponding class
were removed in the simulated events. For $P$ greater than zero, particles 
have to be added to the simulated events.
This was performed copying another particle of the same class and smearing
its momentum by 2.5\% in order not to affect the event jet topology. If there
was no particle of the corresponding class,
a particle of the adjacent class was taken and scaled to fit into this
class.
Note that these modifications of the multiplicities in the Monte Carlo
simulation were not used to change the analysis, but only to estimate the
systematic errors.
The effect on the final background estimation ranges from $\pm$10.5\% (1998), 
$\pm$4.7\%
(1999) to $\pm$10.6\%(2000) for the high mass range analysis.
For the low mass range the effects are smaller, they range from
$\pm$6.6 \% (1998),
$\pm$4.3\% (1999) to $\pm$5.6\% (2000). This procedure also affects the signal
efficiencies leading to
a reduction of the relative signal efficiency of up to $\pm$1.5\%.
The application of this method to the analysis variables leads to a better 
agreement of data at the preselection level as has been observed previously 
in the measurement of the Z pair production cross-section~\cite{ZZ}, leading to 
a better estimation of the systematic error on the simulated background.
The total systematic error and statistical error
from the limited MC statistics are combined in quadrature and given in 
Table~\ref{tab:tabeff} and Table~\ref{tab:tabeff-l}.

%AS100403 MAYBE MOVE TABLE EFF-L HERE

\section{Leptonic channels}

The leptonic channel $\rm H\ell^+\ell^-$ represents about 10\% of the $\rm HZ$ 
final state. The experimental signature of the $\rm HZ(Z \rightarrow \ell^+\ell^-)$
final states is a pair of acoplanar and acollinear leptons, with an
invariant mass compatible with that of an on-shell Z boson.
%the expectation from $\rm Z \rightarrow \ell^+\ell^-$.

The analysis contains a preselection for leptonic events. 
Then, the search channel is defined by the lepton-type of 
the Z decay mode and for each decay mode specific 
selection cuts were applied. 
%Two separated analyses were used 
Two different sets of final cuts were used,
depending on the reconstructed mass, 
defining the low-mass and high-mass ranges.

\subsection{Leptonic preselection}
To ensure a good detector performance the data corresponding to runs in which
subdetectors were not fully operational were discarded. In particular it was
required that the tracking subdetectors and calorimeters were fully 
operational and
%For all the topologies that involve leptons, it was further required 
that the muon chambers were fully functional. 
This resulted in slightly smaller integrated luminosities
than for the hadronic search channel.
An initial set of cuts was applied to select a sample enriched in leptonic
events. A total charged-particle multiplicity 
between 2 and 5 was required. All particles in the event were
clustered into jets using the LUCLUS algorithm~\cite{LUCLUS}
($d_{\rm join}=6.5~\GeVc$) and only events with two reconstructed jets
were retained. Both jets had to contain at least one charged particle
and at least one jet had to contain 
%not more than 
exactly one charged particle.

\begin{figure}[htbp]
\begin{center}
\includegraphics[width=14.0cm]{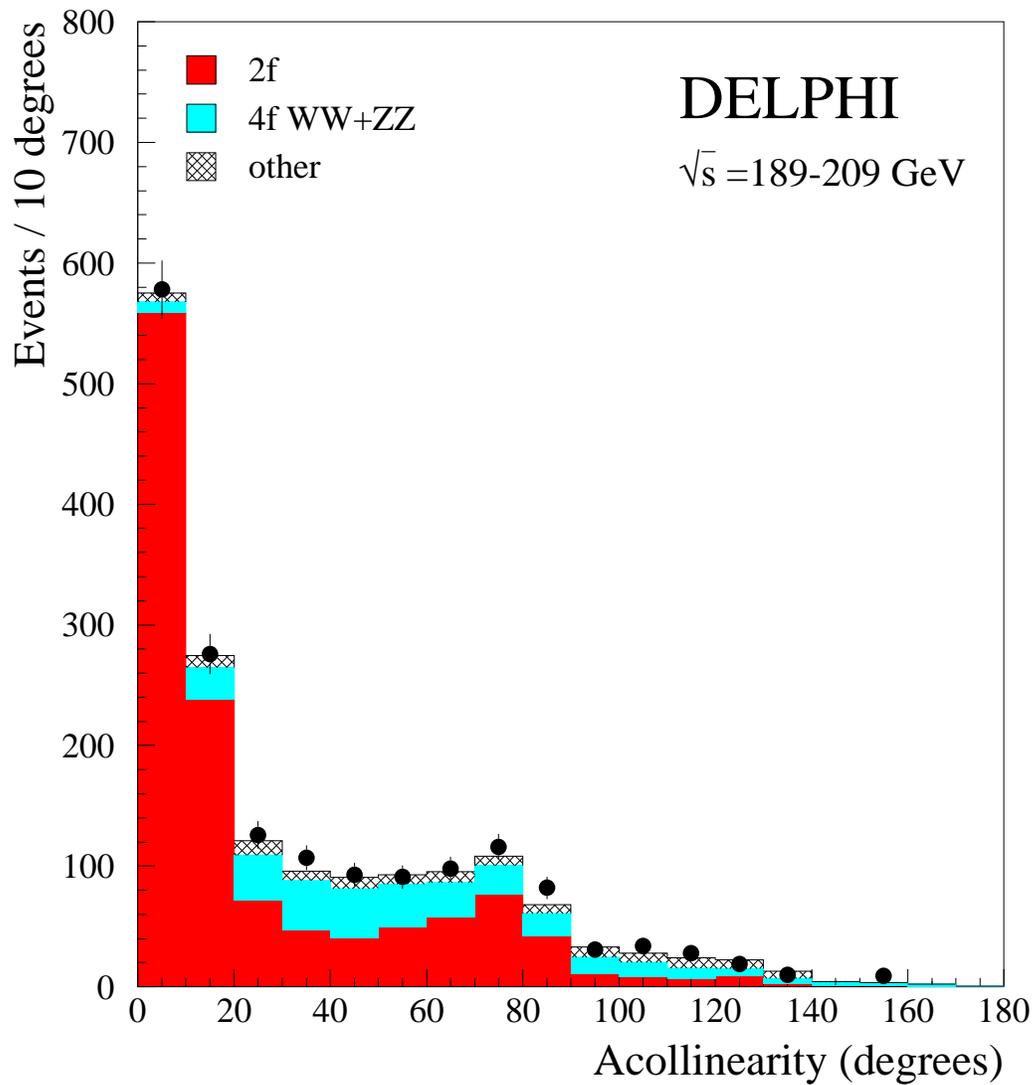} 
\end{center}
\caption{\label{fig:acol}Leptonic channels: Acollinearity distribution 
for $\sqrt{s}$ from 189 to 209~\GeV\
after the preselection.}
\end{figure}

In order to reduce the background from two-photon collisions and 
radiative di-lepton events, the acoplanarity, $\theta_{\rm acop}$, had 
to be larger than 2$^\circ$, and the acollinearity, 
$\theta_{\rm acol}$, had to be larger than 3$^\circ$.
In addition, the total momentum transverse to the beam direction, $p_{\rm T}$,
had to exceed $0.02\sqrt{s}/c$. Finally, the energy of the most energetic 
photon was required to be less than
$0.15\sqrt{s}$. The angle between that photon and the
charged system projected onto the plane perpendicular to the beam axis 
had to be less than 170$^\circ$.
%The number of data and simulated background events are given in
%Table~\ref{tab:tabpresel_ll} for each centre-of-mass energy. 
The agreement of data and background at the preselection level 
is shown in Fig.~\ref{fig:acol} for all data sets.
\begin{table}[htbp]
\begin{center}
\begin{tabular}{|c|c|c|c|c|c|c|}\hline
$\mathrm{\sqrt{s}}$     & \multicolumn{2}{|c|}{$\mathrm{\mu^+\mu^-}$} & \multicolumn{2}{|c|}{$\mathrm{e^+e^-}$} & \multicolumn{2}{|c|}{$\mathrm{\tau^+\tau^-}$}  \\
(GeV)&Data &MC           &Data&MC           &Data &MC            \\\hline
188.6&64   &$49.7\pm0.8$ &314 &$298.0\pm7.1$&124  &$148.2\pm3.7$ \\\hline
191.6&10   &$ 7.9\pm0.1$ & 46 &$ 45.5\pm1.1$& 18  &$ 22.4\pm0.8$ \\\hline
195.5&19   &$22.5\pm0.3$ &132 &$125.5\pm3.1$& 78  &$ 62.2\pm2.1$ \\\hline
199.5&24   &$24.8\pm0.3$ &149 &$134.6\pm3.2$& 81  &$ 74.9\pm1.8$ \\\hline
201.6&17   &$12.1\pm0.2$ & 60 &$ 65.9\pm1.6$& 34  &$ 32.1\pm1.1$ \\\hline
205.0&11   &$20.5\pm0.3$ & 98 &$114.2\pm2.7$& 70  &$ 55.3\pm1.0$ \\\hline
206.5&26   &$23.1\pm0.3$ &110 &$129.5\pm2.2$& 76  &$ 71.9\pm1.1$ \\\hline
206.5U& 6  &$14.6\pm0.2$ & 79 &$ 76.3\pm1.9$& 48  &$ 40.3\pm1.3$ \\\hline
\end{tabular}
\end{center}
\caption{Comparison of simulation and data at preselection level in the three
leptonic channels. The errors reflect the Monte Carlo statistics only.
The last line (206.5U) refers to the data taken
with one TPC sector inoperative, which has been fully taken into
account in the event simulations.
\label{tab:tabpresel_ll}}
\end{table}

\subsection{Channel identification}
For the preselected events, 
jets were then identified as either $\mu$, $e$ or $\tau$ and two
leptons with the same flavour were required.
Owing to the low level of background, the three lepton identifications
rely on loose criteria.
A charged particle was identified as a muon if at least one hit in the
muon chambers was associated to it, or if it had energy deposited in the
outermost layer of the hadron calorimeter. 
%%%A jet was tagged as a muon, if it had only one charged track.
In addition, the energy deposited
in the other layers had to be compatible with that from a minimum ionising
particle.
Only jets with exactly one charged particle were tagged as muons. 
For the identification of a charged 
particle as an electron the energies deposited 
in the electromagnetic calorimeters, in the different layers of the 
hadron calorimeter, and in addition the energy loss in the Time Projection 
Chamber were used.
An electron jet had to contain a maximum of two charged particles with 
at least one identified electron.
A lepton was defined as a cascade decay coming from a $\tau$ 
if the momentum was lower than $0.13\sqrt{s}/c$.
In this case the charged particle is no longer classified as a muon
or as an electron. 
If no muon or electron was identified, the particle was considered 
a hadron from a $\tau$ decay.
Thus, there is no overlap between the event samples selected in the three
channels.
The number of data and simulated background events are given in
Table~\ref{tab:tabpresel_ll} for each centre-of-mass energy.
A detailed description of the lepton identification is given in
Ref.~\cite{wpaper}.

\subsection{Channel-dependent criteria}
After the preselection, different cuts were applied in each channel 
in order to reduce the remaining background. 
The optimisation of the efficiency has been 
performed separately 
for mass ranges of 50 to 85~\GeVcc~ and 85 to 115~\GeVcc.

In the $\rm \mu^+\mu^-$ channel only events with exactly two charged
particle tracks were accepted.
The direction of the missing momentum 
had to deviate from the beam axis by more that $18^\circ$ in order to reject 
$\rm Z\rightarrow\mu^+\mu^-(\gamma)$ and 
$\rm \gamma\gamma\rightarrow\mu^+\mu^-$ processes.
The di-muon mass was required to be between 75~\GeVcc\ 
and 97.5~\GeVcc, to be consistent with the Z boson mass. After that, 
two different sets of cuts were applied depending on the reconstructed 
Higgs boson mass as defined in section~4.5. 
If the reconstructed mass was higher than 85 \GeVcc\ the momentum of the 
most energetic muon had to be between $0.2\sqrt{s}/c$ and $0.4\sqrt{s}/c$.
Furthermore, $E_{\rm vis} < 0.55\sqrt{s}$, $p_{\rm T} < 0.25\sqrt{s}/c$ and 
$\theta_{\rm acol} < 60^\circ$ was required. Otherwise, the momentum of the 
most energetic muon had to be between $0.25\sqrt{s}/c$ and $0.45\sqrt{s}/c$, 
and $0.45\sqrt{s} < E_{\rm vis} < 0.65\sqrt{s}$, 
$p_{\rm T} < 0.4\sqrt{s}/c$ and $45^\circ < \theta_{\rm acol} < 85^\circ$
was required.
The mass resolution for $\rm Z\rightarrow\mu^+\mu^-$ is about 4.5~\GeVcc.

In the $\rm e^+e^-$ channel a maximum of four tracks were required.
The most important background arises 
from radiative Bhabha scattering and $\rm Ze^+e^-$ events. 
To suppress these backgrounds, the direction of the missing momentum
and the polar angle of both leptons had to deviate from the beam axis 
by more than $18^\circ$, the transverse energy had to be greater than 
$0.15\sqrt{s}$ and the neutral electromagnetic energy had to be less 
than $0.1\sqrt{s}$. The invariant mass of the two leptons had to be between 
75~\GeVcc\ and 100~\GeVcc to be consistent with the Z boson mass.
The mass resolution for $\rm Z\rightarrow\ee$ is about 5.7~\GeVcc.
Then, if the mass reconstructed was higher than 85~\GeVcc, the momentum of the 
most energetic electron had to be lower than $0.35\sqrt{s}$, 
and the total associated energy was required to be less than
$0.55\sqrt{s}$, $p_{\rm T} < 0.25\sqrt{s}/c$ and $\theta_{\rm acol} <$ $60^\circ$. 
Otherwise, the momentum of the most energetic electron had to be 
between $0.25\sqrt{s}/c$ and $0.45\sqrt{s}/c$, and 
the total associated energy was required to be less than
$0.65\sqrt{s}$. In addition, the selection $p_{\rm T} < 0.4\sqrt{s}/c$ and 
$45^\circ$ $ < \theta_{\rm acol} < $ $85^\circ$ was applied.

In the $\rm \tau^+\tau^-$ channel tighter cuts were applied on the
acoplanarity and acollinearity in order to reduce the remaining backgrounds
from $\rm \tau^+\tau^-(\gamma)$ and $\rm \gamma\gamma \rightarrow \ell\ell$
processes. 
The invariant mass of both jets had to be less than 3~\GeVcc. 
In addition, the transverse energy had to be greater than $0.1\sqrt{s}$, 
the visible energy of all particles with $\mid \cos\theta \mid<0.9$ 
had to be greater than $0.06\sqrt{s}$ and the energy of both jets had to 
be less than $0.26\sqrt{s}$. Finally, if the mass reconstructed was higher 
than 85~\GeVcc, the acollinearity had to be between 
10$^\circ$ and 60$^\circ$, otherwise, it had to be 
between 45$^\circ$ and 85$^\circ$.
No cut on the reconstructed mass is applied because of the large
missing energy from the associated neutrinos.

\subsection{Systematic uncertainties}
Several sources of systematic uncertainties were investigated
for their effect on the signal efficiency and the background rate. 
The particle identification method was checked with di-lepton samples both 
at Z peak and high energy, and the simulation and data rates were found 
to agree within $\pm1\%$.
%The errors on the background and signal rates from 
The modelling of the preselection variables agrees within statistical 
errors with the data.
%and the detector response were a few percent. 
The track selection and the track reconstruction efficiency were also
taken into account in the total systematic error. 
The effects of detector miscalibration and deficiencies were 
investigated using 
%events compatible with 
$\mu^+\mu^-\gamma$ or $e^+e^-\gamma$ events, where the lepton energies
are determined directly and recoiling from the photon.
The comparison between data and simulation rate was found to be better 
%%%AS: too precise! than 0.9\%. The finite size of
than $\pm1\%$.
%%%AS THIS IS STATISTICS: The finite size of the Monte Carlo samples 
%%% used in the analysis results an additional uncertainty
%%% for the signal efficiency and the background. 
Additional systematic effects were estimated by comparing the data 
collected, at the Z peak, during the period with one TPC sector inoperative 
with simulation samples produced with the same detectors conditions.
%AS ERROR: The additional systematic
The total systematic error on the signal efficiency was $\pm 1.1\%$.
The total systematic error on the background rate was up to $10\%$.
The total systematic error and statistical error
from the limited MC statistics are combined in quadrature and given in 
Table~\ref{tab:tabeff}.

\subsection{Mass reconstruction}
The mass of the invisibly decaying particle was computed from the measured
energies assuming momentum and energy conservation. To improve the
resolution a $\chi^2$ fit was applied constraining the visible mass 
to be compatible with a Z. 
In the case of the $\tau^+\tau^-$ channel, the 
%information carried by  
measured four-momenta of 
the decay products 
do not reproduce correctly the $\tau$ energy. 
Therefore, the mass was calculated under the assumption that both $\tau$ 
leptons had the same energy and the $\tau$ neutrino went along the 
direction of the $\tau$ lepton.
This, together with the visible mass constraint, 
allowed an estimation of the $\tau$ energy and of the invisible mass. 
The invisible mass for the candidates as well as for the expected 
background from Standard Model processes for the different
channels is shown in Fig.~\ref{fig:maslep}.

\section{Results}
A comparison of the observed and predicted numbers of selected events
for the four channels is summarised in Tables \ref{tab:tabeff} and \ref{tab:tabeff-l}.
The agreement between the data and the SM prediction  is good for all channels and
no indication for a Higgs boson decaying into invisible particles has been observed.
The signal efficiencies of the four channels are shown in Fig. \ref{pic:eff} 
as a function of the Higgs mass for $\sqrt{s}=206.5$~GeV.

\begin{table}[htbp]
\begin{center}
\begin{tabular}{|c|c|r|r|c|c|}
\hline
$\mathrm{\sqrt{s}}$ & Channel & Luminosity & Data & Expected & Signal efficiency \\
(GeV) & \mbox{ } & (pb$^{-1})$ & \mbox{ } & background &  (\%) \\
\hline
188.6   & $\mathrm{q\bar{q}}$   &152.4  & 65 &$71.3\pm7.7$ &$40.9\pm1.9$\\
191.1   & $\mathrm{q\bar{q}}$   & 24.7  &  2 &$ 5.6\pm0.3$ &$39.6\pm1.7$\\
195.5   & $\mathrm{q\bar{q}}$   & 74.3  & 21 &$18.7\pm1.0$ &$50.8\pm1.7$\\
199.5   & $\mathrm{q\bar{q}}$   & 82.2  & 21 &$20.1\pm1.0$ &$51.9\pm1.7$\\
201.6   & $\mathrm{q\bar{q}}$   & 40.0  & 11 &$10.8\pm0.5$ &$50.7\pm1.7$\\
205.0   & $\mathrm{q\bar{q}}$   & 74.3  &  9 &$12.2\pm1.3$ &$36.4\pm2.1$\\
206.5   & $\mathrm{q\bar{q}}$   & 82.8  & 13 &$13.5\pm1.5$ &$37.0\pm2.1$\\
206.5U  & $\mathrm{q\bar{q}}$   & 58.0  & 11 &$ 8.4\pm0.9$ &$31.6\pm2.1$\\ 
\hline
188.6   & $\mathrm{\mu^+\mu^-}$ &153.8  &  7 &$ 6.9\pm0.6$ &$44.0\pm1.9$ \\
191.1   & $\mathrm{\mu^+\mu^-}$ & 24.5  &  4 &$ 1.1\pm0.1$ &$52.8\pm1.6$ \\
195.5   & $\mathrm{\mu^+\mu^-}$ & 72.4  &  3 &$ 3.5\pm0.2$ &$63.8\pm1.5$ \\
199.5   & $\mathrm{\mu^+\mu^-}$ & 81.8  &  0 &$ 3.9\pm0.3$ &$63.0\pm1.5$ \\
201.6   & $\mathrm{\mu^+\mu^-}$ & 39.4  &  2 &$ 1.8\pm0.2$ &$62.5\pm1.5$ \\
205.0   & $\mathrm{\mu^+\mu^-}$ & 69.1  &  0 &$ 3.0\pm0.3$ &$62.8\pm1.5$ \\
206.5   & $\mathrm{\mu^+\mu^-}$ & 79.8  &  2 &$ 3.3\pm0.3$ &$62.1\pm1.5$ \\
206.5U  & $\mathrm{\mu^+\mu^-}$ & 50.0  &  0 &$ 2.2\pm0.2$ &$56.9\pm1.6$ \\
\hline
188.6   & $\mathrm{e^+e^-}$     &153.8  & 4  &$ 7.9\pm0.7$ &$34.2\pm1.3$ \\
191.1   & $\mathrm{e^+e^-}$     & 24.5  & 1  &$ 1.2\pm0.2$ &$40.8\pm1.6$ \\
195.5   & $\mathrm{e^+e^-}$     & 72.4  & 4  &$ 4.7\pm0.5$ &$45.3\pm1.6$ \\
199.5   & $\mathrm{e^+e^-}$     & 81.8  & 5  &$ 4.1\pm0.4$ &$45.2\pm1.6$ \\
201.6   & $\mathrm{e^+e^-}$     & 39.4  & 1  &$ 1.9\pm0.2$ &$45.1\pm1.6$ \\
205.0   & $\mathrm{e^+e^-}$     & 69.1  & 3  &$ 3.6\pm0.3$ &$44.8\pm1.6$ \\
206.5   & $\mathrm{e^+e^-}$     & 79.8  & 1  &$ 4.0\pm0.4$ &$42.9\pm1.6$ \\
206.5U  & $\mathrm{e^+e^-}$     & 50.0  & 1  &$ 2.3\pm0.3$ &$39.9\pm1.6$ \\
\hline
188.6   & $\mathrm{\tau^+\tau^-}$ &153.8  & 7  &$ 9.4\pm0.8$ &$ 21.4\pm1.4$ \\
191.1   & $\mathrm{\tau^+\tau^-}$ & 24.5  & 1  &$ 1.9\pm0.2$ &$ 17.3\pm1.4$ \\
195.5   & $\mathrm{\tau^+\tau^-}$ & 72.4  & 7  &$ 5.7\pm0.6$ &$ 20.2\pm2.1$ \\
199.5   & $\mathrm{\tau^+\tau^-}$ & 81.8  &10  &$ 6.3\pm0.6$ &$ 27.3\pm1.5$ \\
201.6   & $\mathrm{\tau^+\tau^-}$ & 39.4  & 2  &$ 3.3\pm0.4$ &$ 28.2\pm1.5$ \\
205.0   & $\mathrm{\tau^+\tau^-}$ & 69.1  & 5  &$ 5.7\pm0.6$ &$ 29.5\pm1.5$ \\
206.5   & $\mathrm{\tau^+\tau^-}$ & 79.8  & 3  &$ 7.1\pm0.7$ &$ 30.3\pm1.5$ \\
206.5U  & $\mathrm{\tau^+\tau^-}$ & 50.0  & 2  &$ 4.5\pm0.4$ &$ 29.5\pm1.5$ \\\hline
\end{tabular}                                                                  
\end{center}
\caption[]{Integrated luminosity, observed number of events, 
expected number of background events and signal efficiency 
(100~\GeVcc\ signal mass) for different energies. The last
lines of each channel (206.5U) refers to the data taken with one TPC sector 
inoperative, which has been fully taken into account in the 
event simulations. 
Systematic and statistical errors are combined in quadrature from the
results of each analysis.}
\label{tab:tabeff}
\end{table}

\begin{table}[h]
\begin{center}
\begin{tabular}{|c|c|r|r|c|c|}
\hline
$\mathrm{\sqrt{s}}$&Channel&Luminosity&Data&Expected&Signal efficiency \\
(GeV) & \mbox{ } & (pb$^{-1})$ & \mbox{ } & Background &  (\%) \\\hline
188.6 & $\mathrm{q\bar{q}}$   &152.4  & 58 &$51.5\pm3.8$  &$49.1\pm1.6$\\
191.6 & $\mathrm{q\bar{q}}$   & 24.7  & 6  &$10.1\pm0.5$  &$50.0\pm1.7$\\
195.5 & $\mathrm{q\bar{q}}$   & 74.3  & 36 &$31.3\pm1.6$  &$49.6\pm1.7$\\
199.5 & $\mathrm{q\bar{q}}$   & 82.2  & 37 &$44.3\pm2.3$  &$50.5\pm1.7$\\
201.6 & $\mathrm{q\bar{q}}$   & 40.0  & 10 &$12.0\pm0.6$  &$44.2\pm1.7$\\
205.0 & $\mathrm{q\bar{q}}$   & 74.3  & 26 &$26.2\pm1.7$  &$47.0\pm1.5$\\
206.5 & $\mathrm{q\bar{q}}$   & 82.8  & 30 &$33.4\pm2.1$  &$48.8\pm1.5$\\
206.5U& $\mathrm{q\bar{q}}$   & 58.0  & 10 &$18.0\pm1.2$  &$43.6\pm1.5$\\
\hline
\end{tabular}                                                                  
\end{center}
\caption[]{Integrated luminosity, observed number of events, 
expected number of background events and signal efficiency 
(60~\GeVcc\ signal mass) for different energies in the low mass analysis. 
The last lines of each channel (206.5U) refers to the data taken with 
one TPC sector inoperative, which has been fully taken into account in the 
event simulations.
Systematic and statistical errors are combined in quadrature.}
\label{tab:tabeff-l}
\end{table}

\subsection{Model independent limits} 
\label{sec:ind}
The cross-section and mass limits were computed at the 95\% CL with a
likelihood method~\cite{ALRMC}.
One-dimensional distributions of the reconstructed mass serve as input
for the likelihood calculation.
The impact of the correlation of the systematic errors is small 
and the limits result largely from the data taken at the higher
centre-of-mass energies. 
More details about the confidence definition and computation
can be found in Ref.~\cite{DELPHIh2000}.
All search channels and centre-of-mass energies were treated as 
separate experiments to obtain
a likelihood function. In total 40 channels were evaluated 
as listed in Tables~\ref{tab:tabeff} and~\ref{tab:tabeff-l}, in addition
to the $\mathrm{q\bar{q}}$ channels from 161 and 172~\GeV\
data~\cite{DELPHIh96}, and the $\mathrm{q\bar{q}}$ and
$\mathrm{\mu^+\mu^-}$ channels from 183~\GeV\ 
data~\cite{DELPHIi97}.
In order to address the overlap between the low and high mass analyses
in the hadronic channel, the expected performance was calculated 
for both analyses in the overlap region. 
%The analysis being more performant at a given mass, 
At each test mass the analysis with the best expected exclusion
power was then chosen for the calculation of the limit. 
 
\begin{figure}[htbp]
\begin{center}
\subfigure[]{
\includegraphics[width=7.5cm]{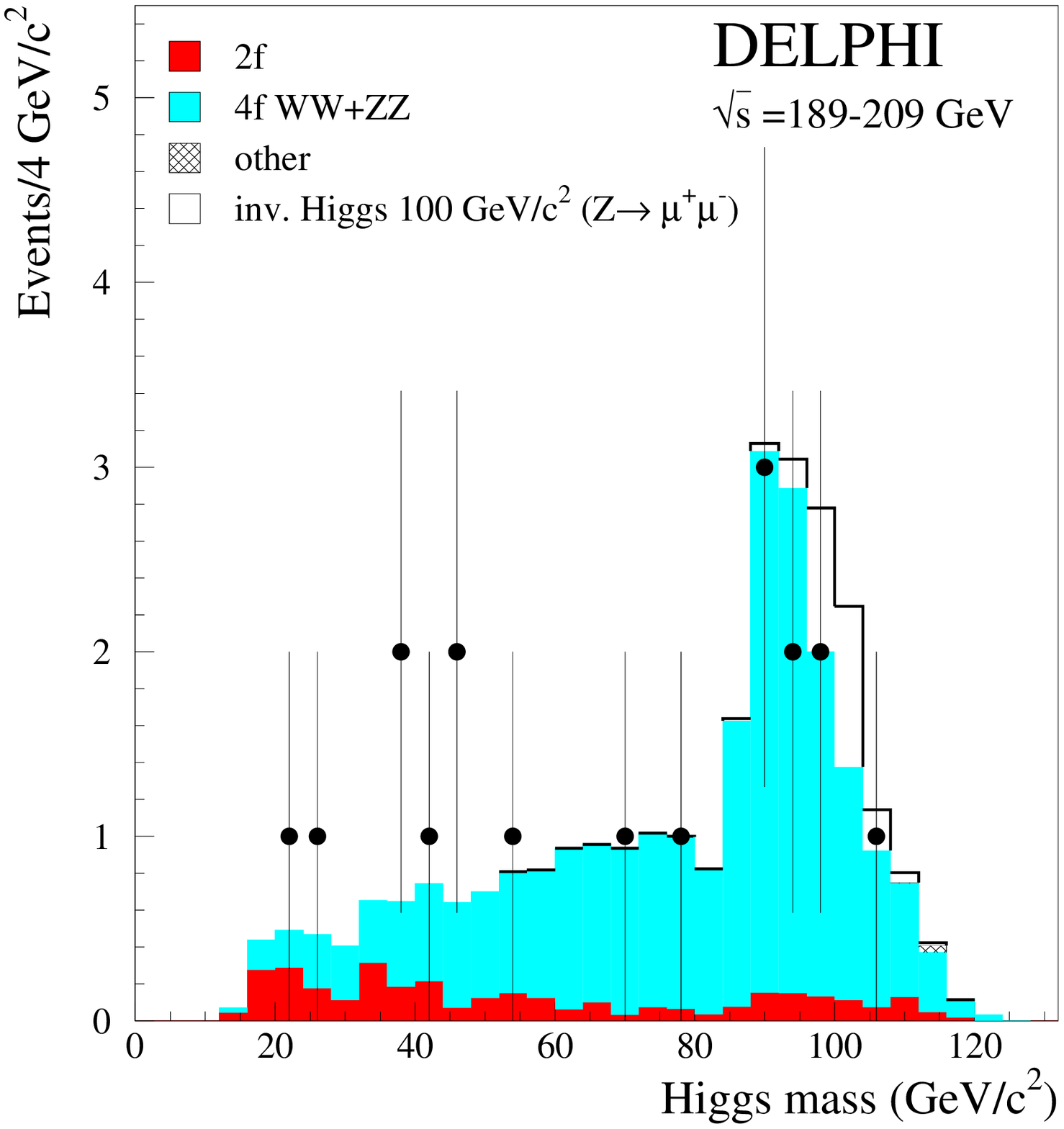}
}
\subfigure[]{
\includegraphics[width=7.5cm]{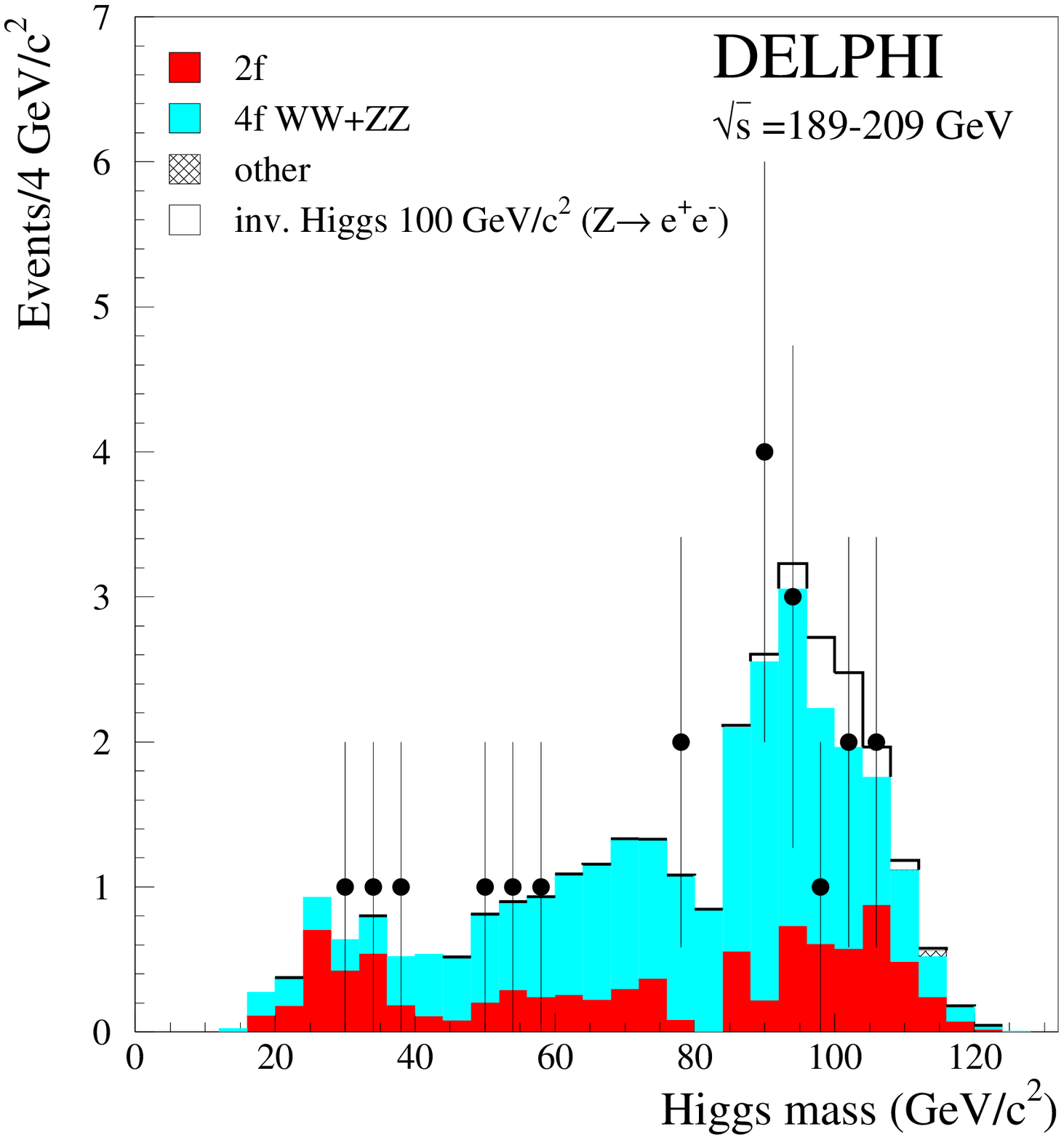}
}
\subfigure[]{
\includegraphics[width=7.5cm]{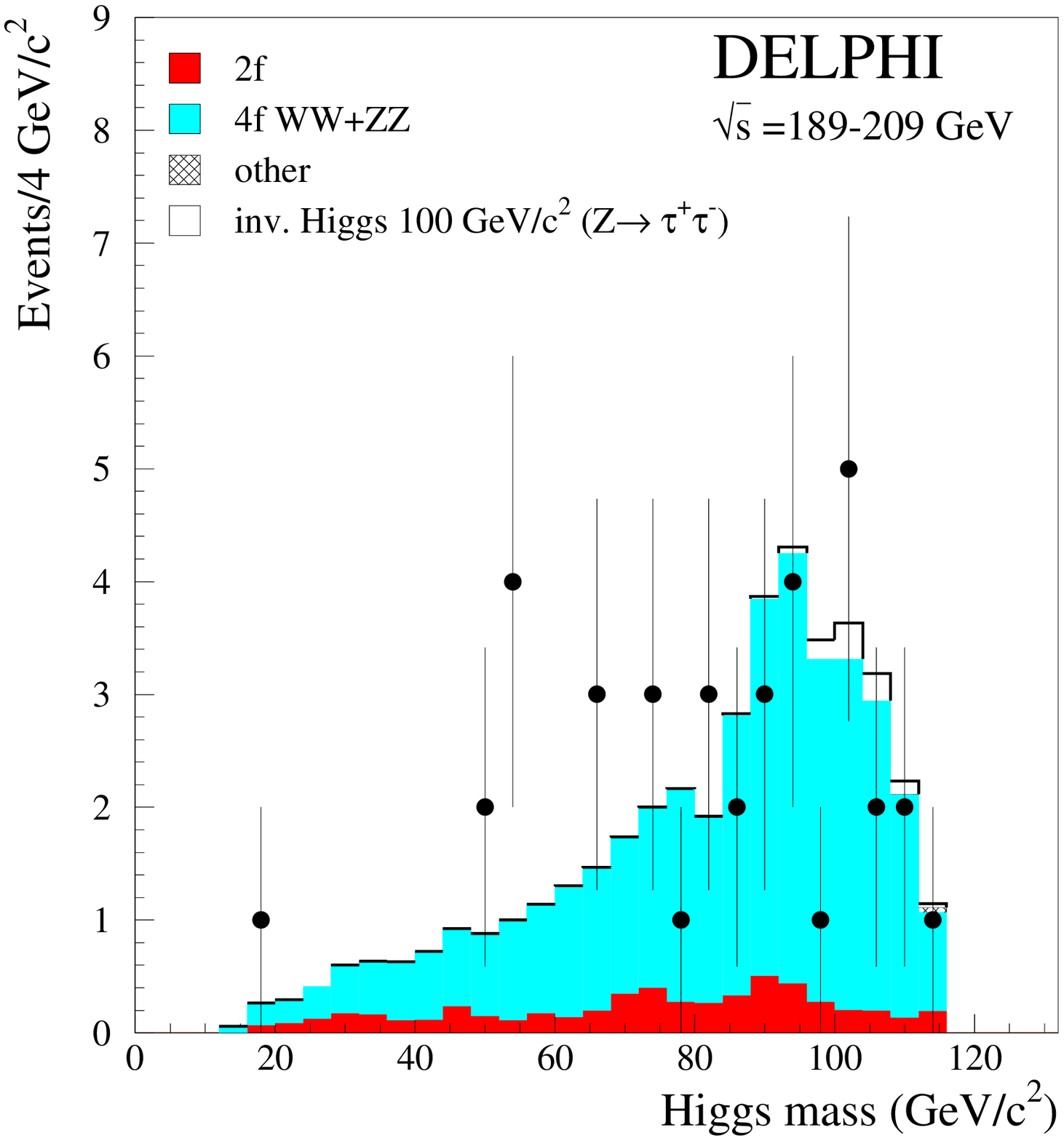}
}
\end{center}
\caption{\label{fig:maslep}Leptonic channels: Reconstructed Higgs boson mass
in (a) the \hmumu channel, (b) the \hee channel and (c) the \htautau channel for 189 to 209~\GeV\
after the final selection. The white histogram corresponds to a Higgs boson with 100~\GeVcc\ mass decaying 
to 100\% into invisible modes.}
\end{figure}

\begin{figure}[htbp]
\begin{center}
\includegraphics[width=15cm]{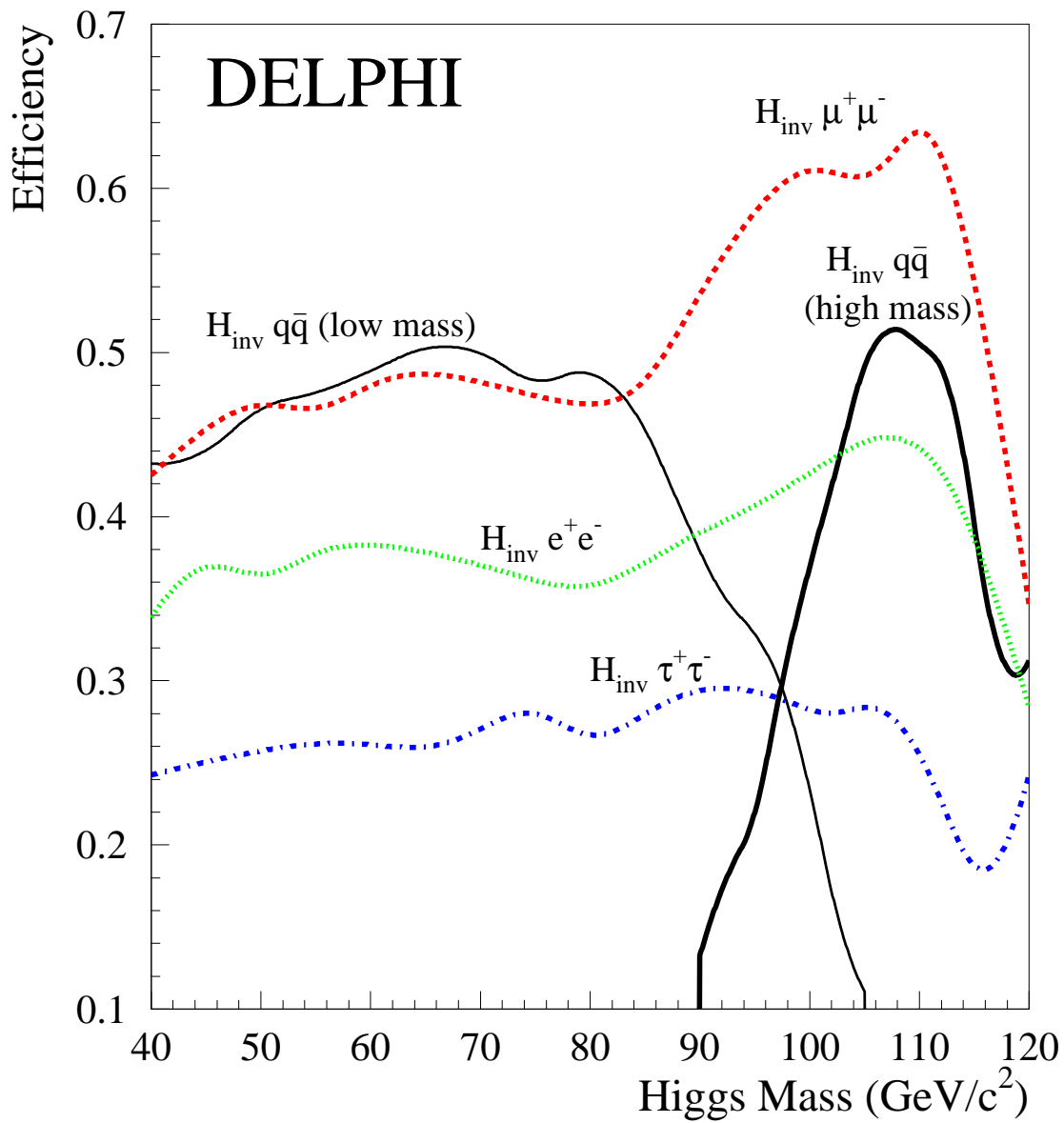}
\caption{\label{pic:eff}
Efficiencies for the Higgs boson masses between 40 and 120 \GeVcc\
for the different selection channels at $\sqrt{s}=206.5$~GeV.}
\end{center}
\end{figure}

\begin{figure}[htbp]
\begin{center}
\subfigure[]{
\includegraphics[width=7.5cm]{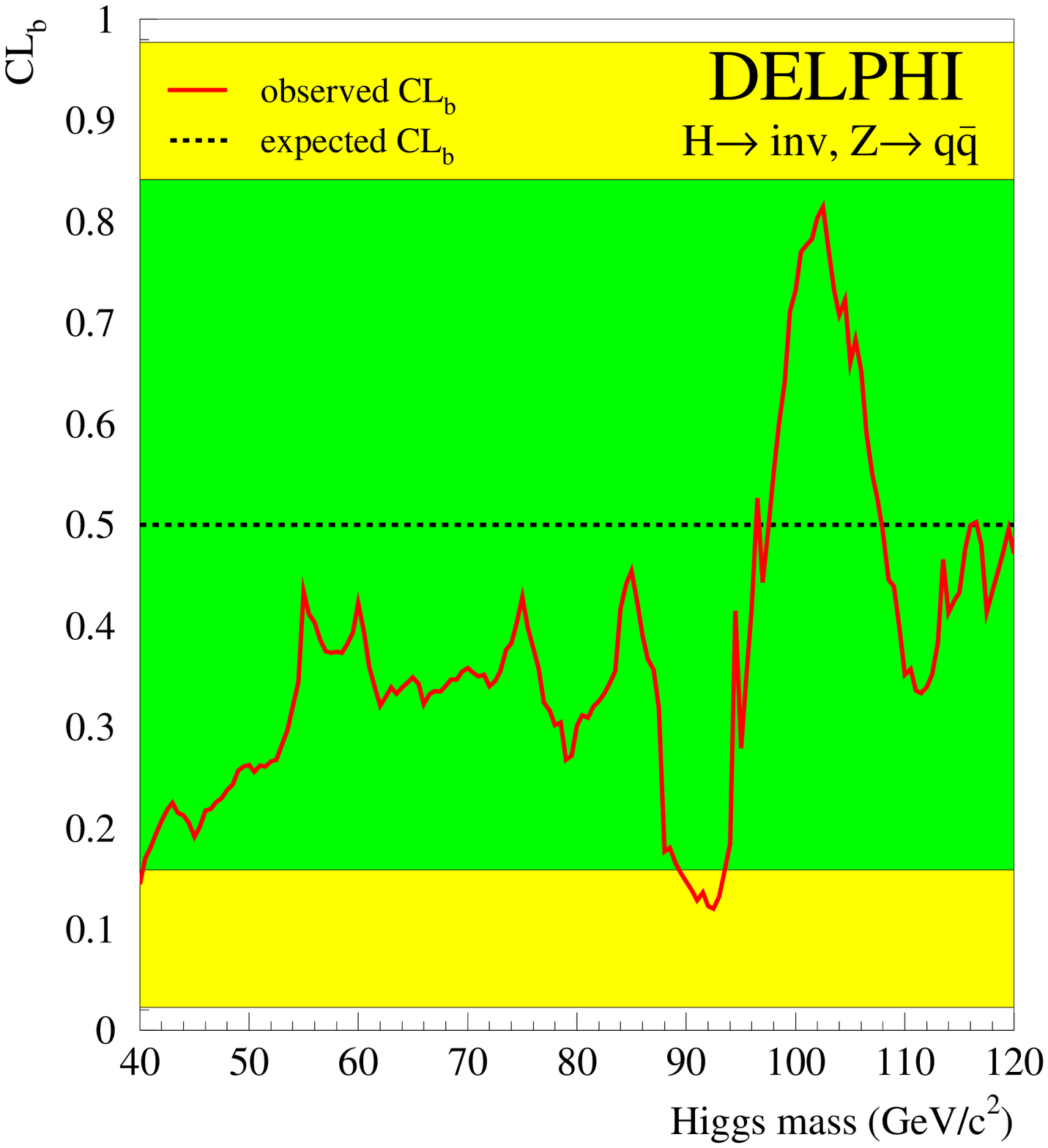}
}
\subfigure[]{
\includegraphics[width=7.5cm]{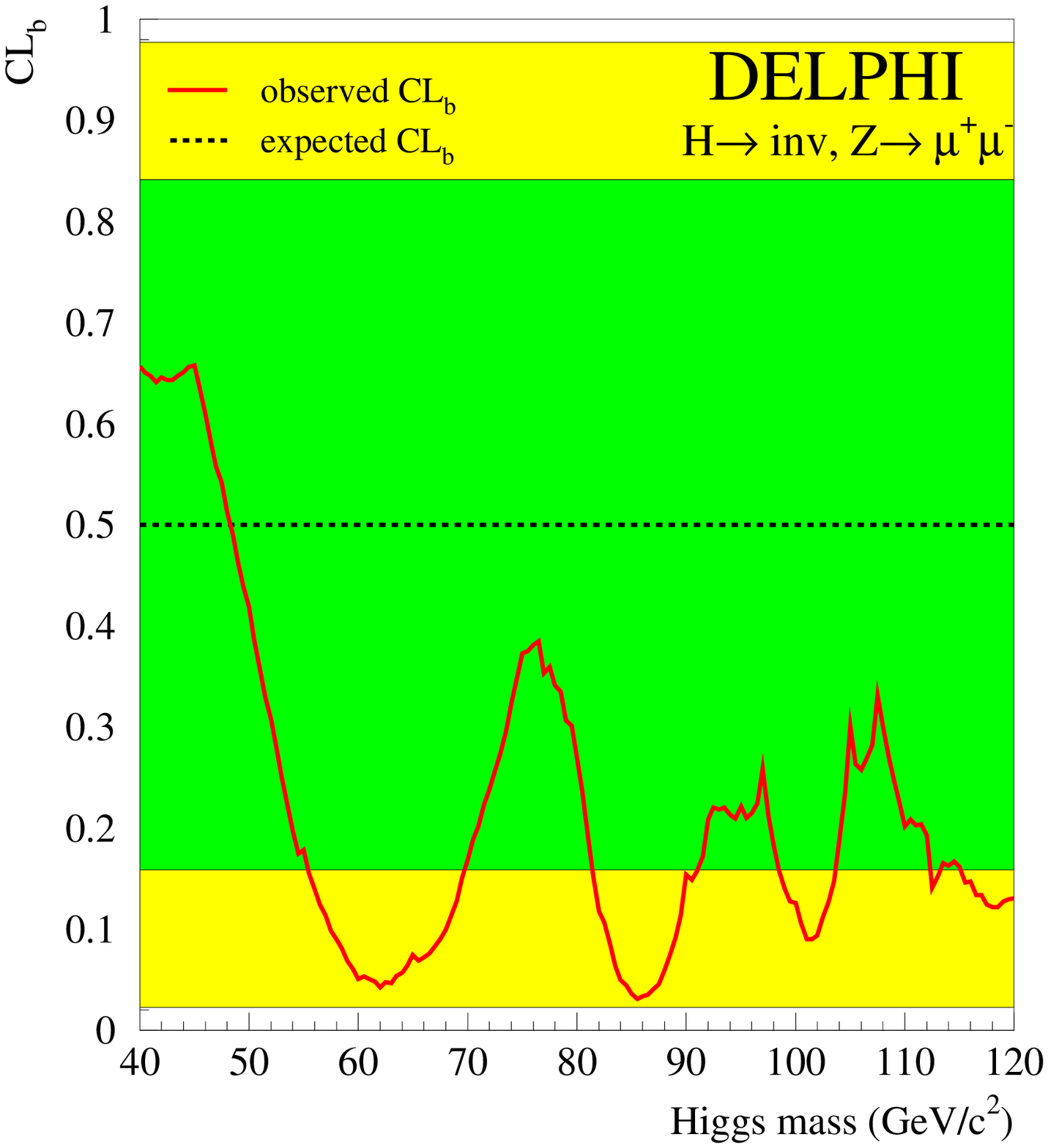}
}
\subfigure[]{
\includegraphics[width=7.5cm]{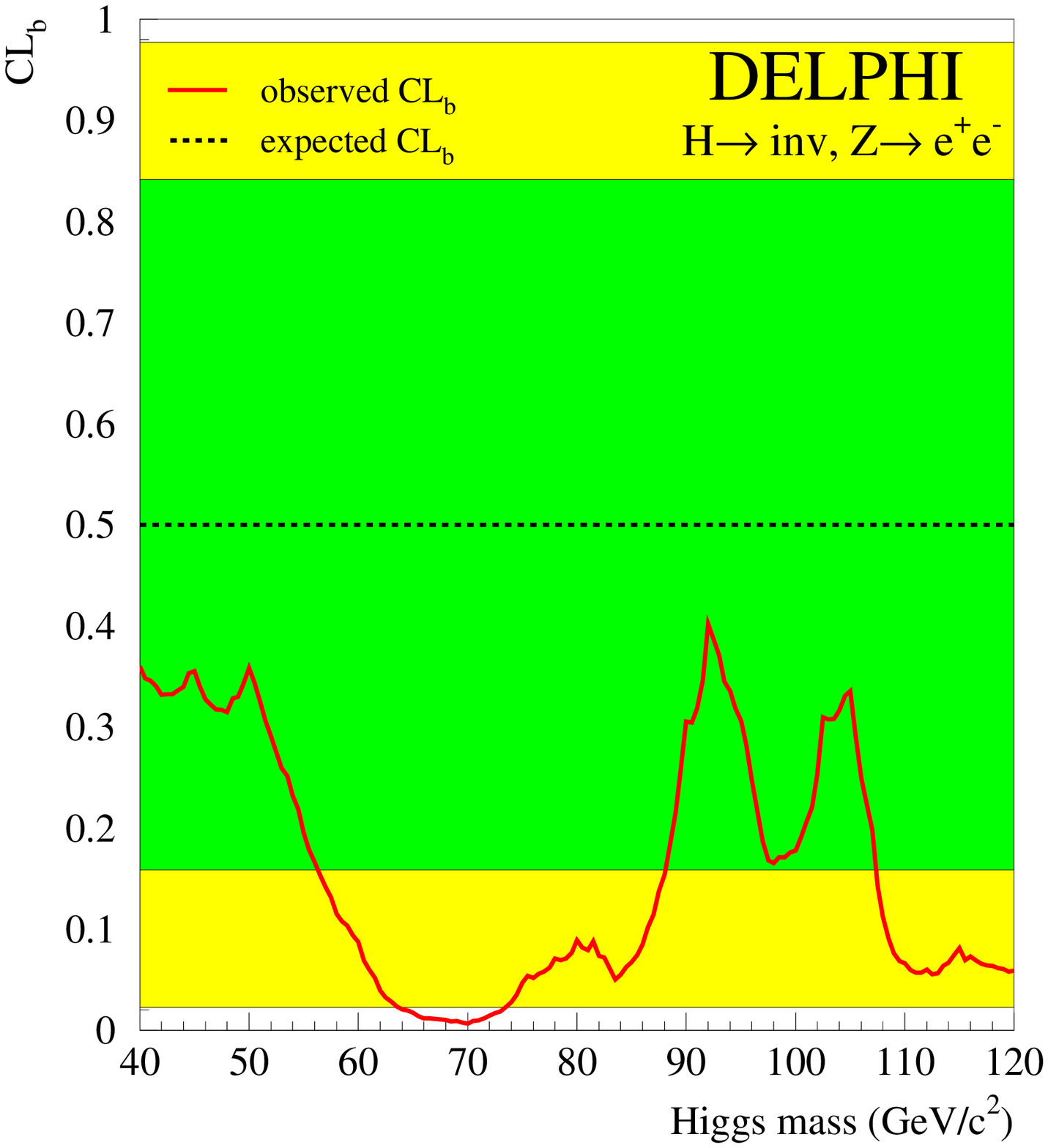}
}
\subfigure[]{
\includegraphics[width=7.5cm]{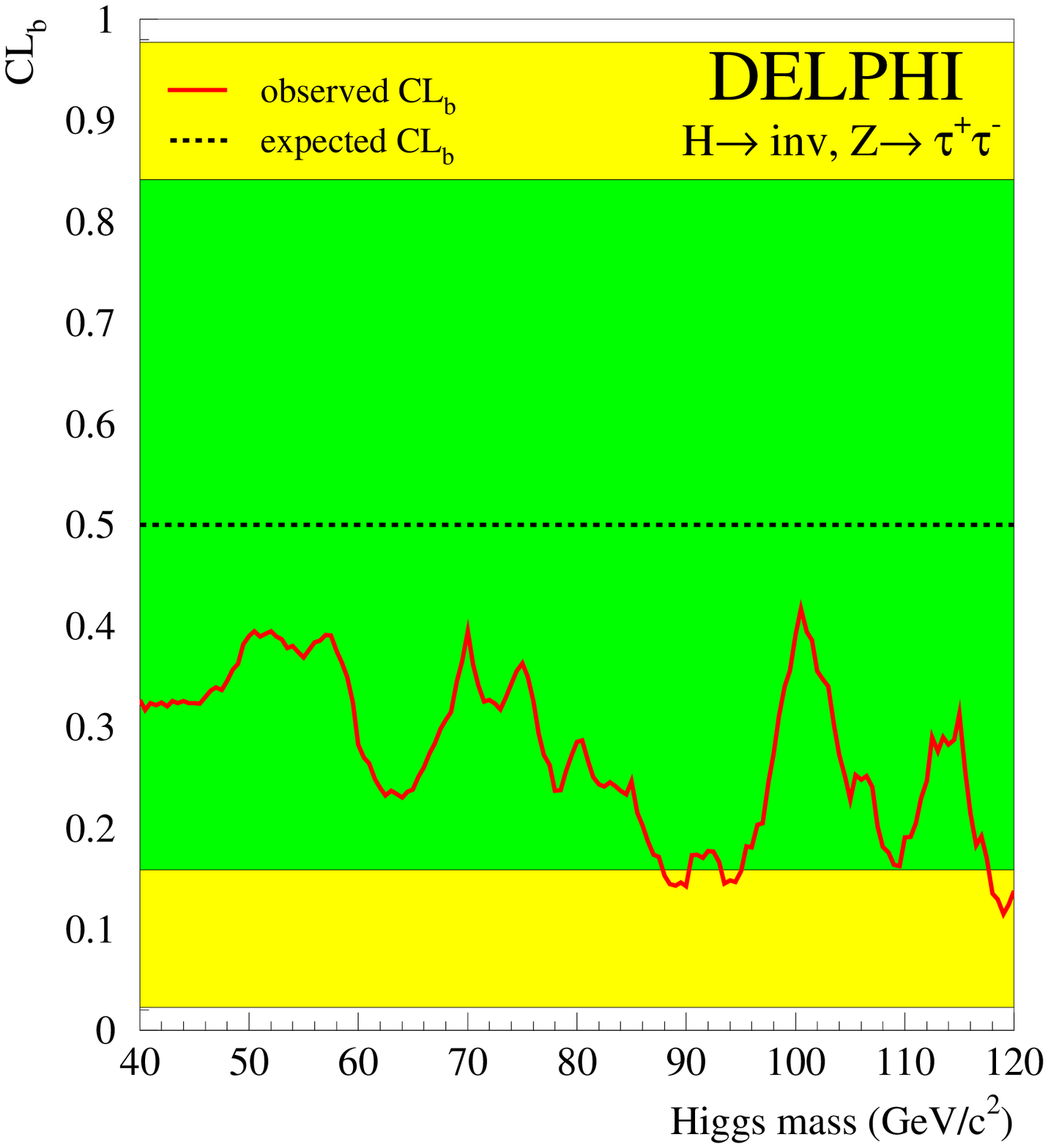}
}
\end{center}
\caption{\label{fig:clb-plots}
Confidence levels for the different decay channels as a function of the Higgs mass. Shown are the 
observed (solid) and expected  (dashed) confidences for the background-only 
hypothesis in the \hqq (a), \hmumu (b),
 \hee (c) and \htautau (d) channels. The dark grey band corresponds
to the 68.3\% expected confidence interval and the light grey band to the 95.0\% confidence interval.
The structures near 94 and 96~GeV in plot (a) are due to the switching from the low-mass to
the high-mass optimization in the hadronic channel.}
\end{figure}

No indication of a signal is observed above the background expectation.
This is shown in Fig.~\ref{fig:clb-plots}
which displays the curves of the confidence levels in the
background hypothesis, $\rm CL_b$, as a function of the Higgs boson mass
hypothesis, for each channel separately. Over most of the range of masses
the agreement between data and the background expectations is within one
standard deviation. However, at a few masses in the muon and electron
channels, there are disagreements near or slightly above two standard
deviations, which are due to deficits of data in several bins of the
reconstructed mass spectra in these channels, as shown in Fig.~\ref{fig:maslep}.
%The structures in the $CL_{\rm b}$ distributions correspond to the distribution 
%of candidates as shown in Figs.~\ref{fig:massqq},~\ref{fig:massqq-l} 
%and~\ref{fig:maslep}.
Figure \ref{pic:lim_inv} displays the 
observed and expected upper limits on the cross-section for the process 
$ \ee \rightarrow $~Z(anything)H(invisible) as a function of the Higgs boson 
mass.
From the comparison with the Standard Model (SM) Higgs boson cross-section 
the observed (expected median) mass limits are 112.1 (110.5)~\GeVcc\
for the Higgs boson decaying into invisible particles.

In a model-independent approach the branching ratio into invisible 
particles $BR_{\rm inv}$ can be considered a free parameter. 
The remaining decay modes are then visible 
and are assumed to follow the SM decay probabilities. 
In this case the searches for visible and invisible Higgs boson decays 
can be combined to determine the excluded region in the $BR_{\rm inv}$ 
versus $m_H$ plane assuming SM production cross-sections. 
Using the DELPHI data from the SM Higgs searches
\cite{DELPHIh96,DELPHIh2000,DELPHIh97,DELPHIh98,DELPHIh99}
a lower mass limit of 111.8~\GeVcc~can be set independently of the
hypothesis on the fraction of invisible decay modes, as shown in
Fig.~\ref{pic:lim_comb}.
In computing these limits, the overlap between the standard $\hnn$ 
and the invisible Higgs boson hadronic selections have been avoided,
conservatively for the limit, 
by omitting the $\hnn$ ($\rm H_{inv}q\bar{q}$) results in the 
region $BR_{\rm inv} >50\% (<50\%)$.

\subsection{Limits for a Majoron model}
The limits computed above can be used to set a limit on the Higgs 
bosons in a Majoron model~\cite{peccei,campos,wells} 
with one complex doublet $\phi$ and one complex singlet $\eta$.
Mixing of the real parts of $\phi$ and $\eta$ leads to two massive Higgs 
bosons:
\begin{displaymath}
H=\phi_R\cos\theta - \eta_R \sin\theta
\end{displaymath}
\begin{displaymath}
S=\phi_R\sin\theta + \eta_R \cos\theta
\end{displaymath}
where $\theta$ is the mixing angle. The imaginary part of the singlet 
is identified as the Majoron. 
The Majoron is decoupled from the fermions and gauge bosons, but
might have a large coupling to the Higgs bosons. 
In this model the free parameters are the masses of H and S, 
the mixing angle $\theta$ and the ratio of the vacuum expectation
values of the two fields $\phi$ and $\eta$ 
($\tan\beta \equiv \frac{v_\phi}{v_\eta}$).
The production rates of the H and S are reduced with respect to the SM Higgs
boson, by factors of $\cos^2\theta$\ and $\sin^2\theta$, respectively.
The decay widths of the H and S into the heaviest possible fermion-antifermion
pair are reduced by the same factor and their decay widths into a Majoron pair
are proportional to the complementary factors ($\cos^2\theta$\ for S and
$\sin^2\theta$\ for H).
The HZ and SZ cross-section times branching ratio into 
invisible decays is calculated and compared to the excluded cross-section of 
section~\ref{sec:ind}.
In the case where the invisible Higgs boson
decay mode is dominant ($\tan \beta$ larger than about 10),
the excluded region in the mixing angle versus Higgs boson mass plane
is shown in Fig.~\ref{pic:majoron}.

\subsection{Limits in the MSSM}
\begin{figure}[htbp]
\begin{center}
\epsfig{file=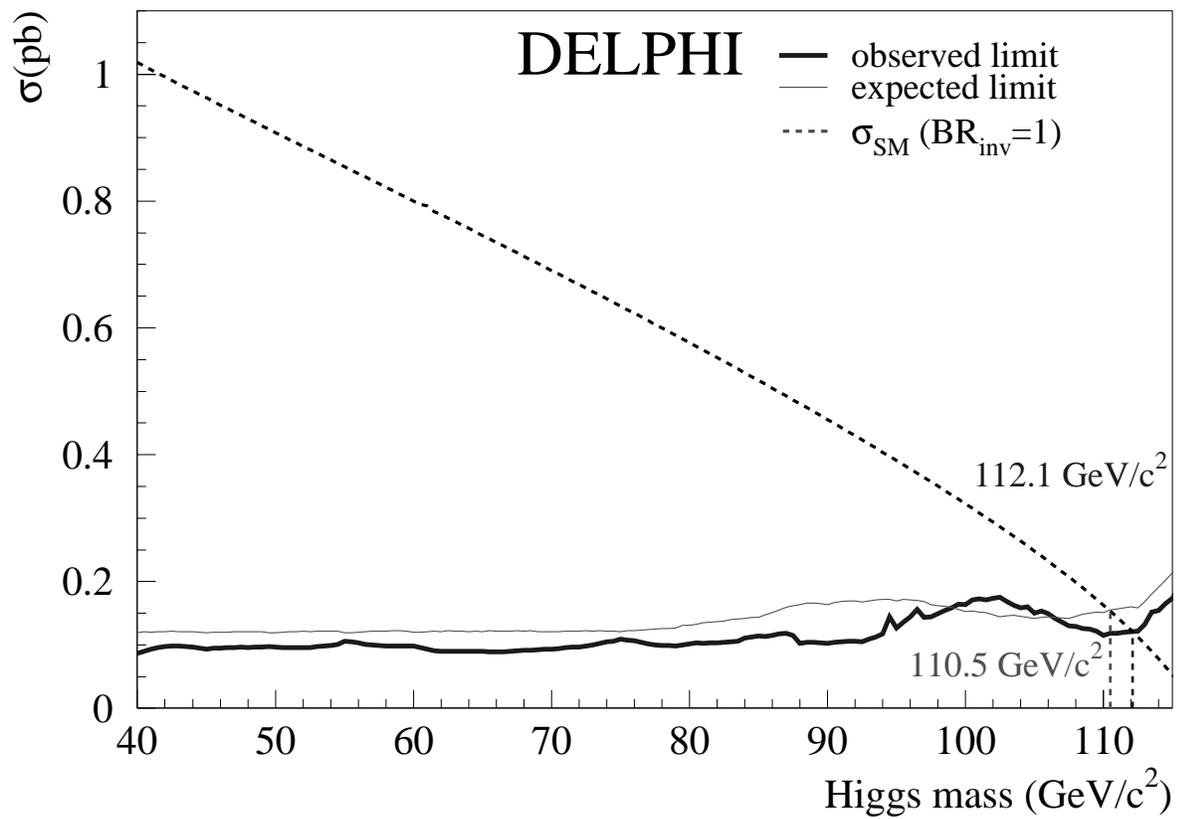,width=16 cm}
\caption[]{
The 95\% CL upper limit on the cross-section 
$\rm \ee \rightarrow Z(\mbox{anything}) \, H(\mbox{invisible})$ as a function 
of the Higgs boson mass. The dashed line shows the standard model 
cross-section for the Higgs boson production with $BR_{\rm inv}= 1$.}
\label{pic:lim_inv}
\end{center}
\end{figure}

\begin{figure}[htbp]
\begin{center}
\epsfig{file=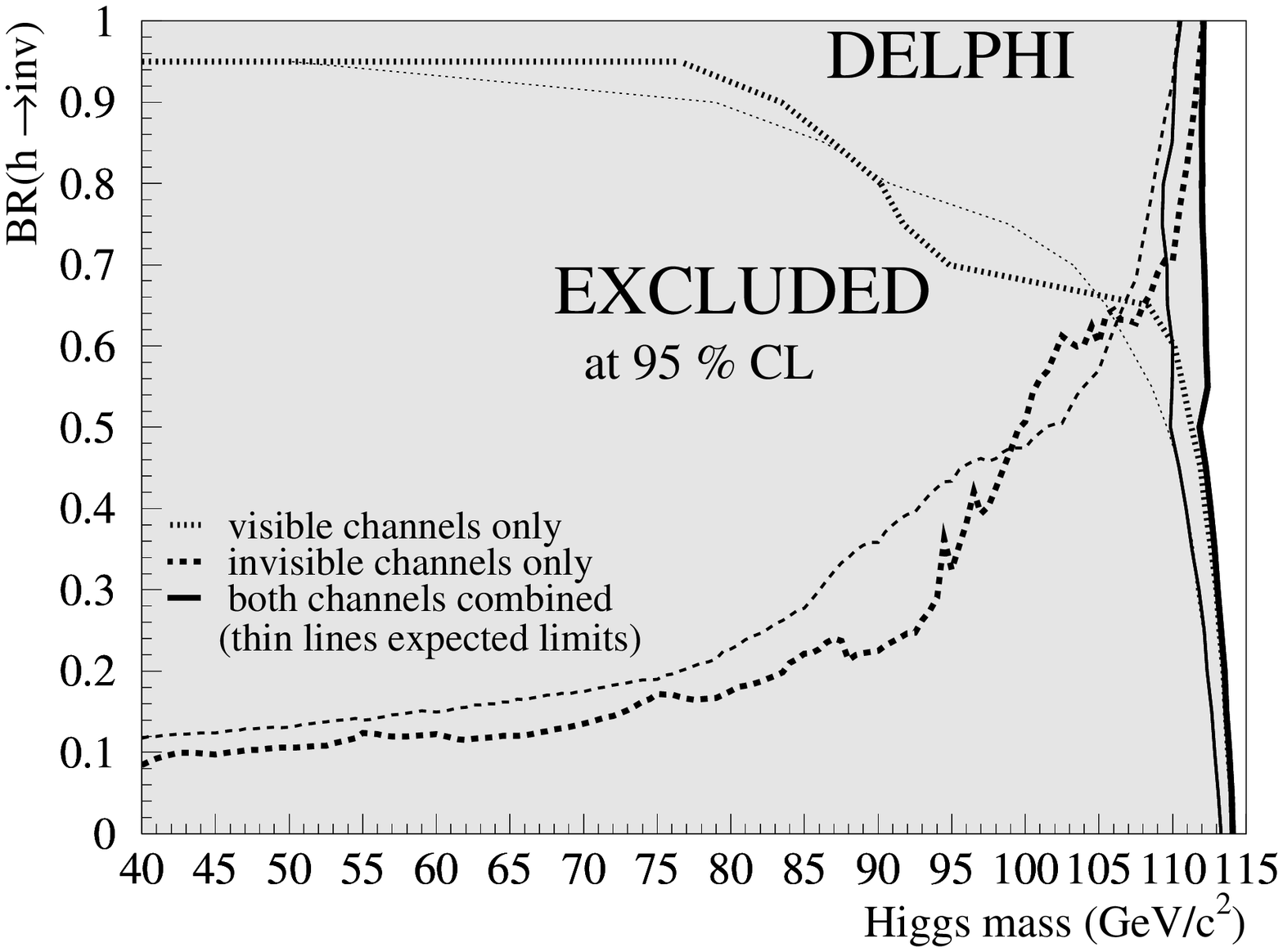 ,width=16 cm}
\caption[]{%\hqq\ channel: 
The Higgs boson mass limits as a function of the branching ratio 
into invisible decays $BR_{\rm inv}$, assuming a 
$1-BR_{\rm inv}$ branching ratio into standard visible decay modes. 
}
\label{pic:lim_comb}
\end{center}
\end{figure}

\begin{figure}[htbp]
\begin{center}

\epsfig{file=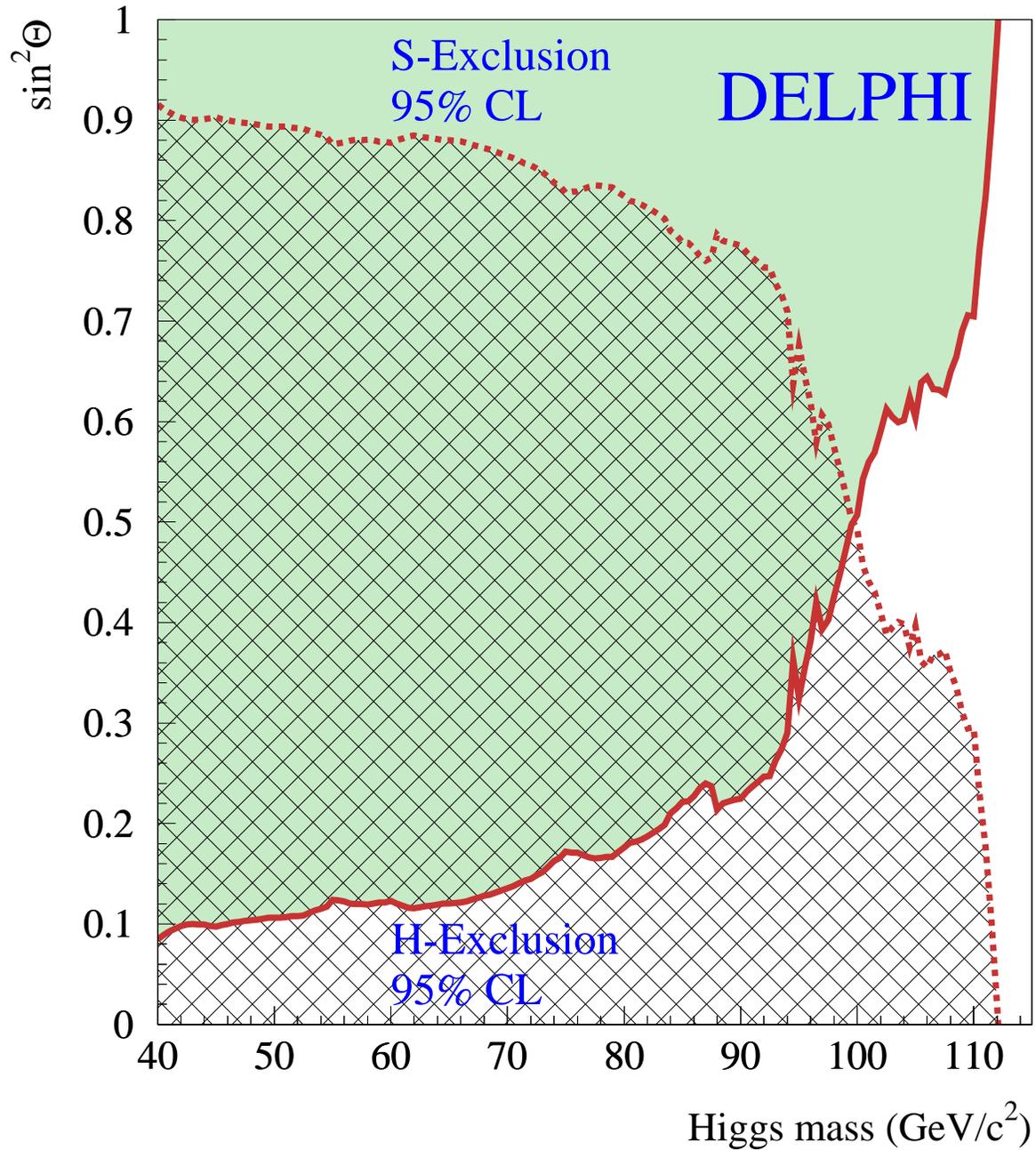,width=16 cm}
\caption[]{%\hqq\ channel: 
Limit on $\sin^2\theta$ as a function of the Higgs boson mass at 95\% CL. 
S and H are the Higgs bosons in the Majoron model. The grey region is excluded
for the S Higgs boson and the hatched region for the H Higgs boson.
The massive Higgs bosons decay almost entirely into invisible Majoron pairs
for large $\tan\beta$ values. 
}
\label{pic:majoron}
\end{center}

\end{figure}

\begin{figure}[htbp]
\begin{center}
\includegraphics[width=15cm]{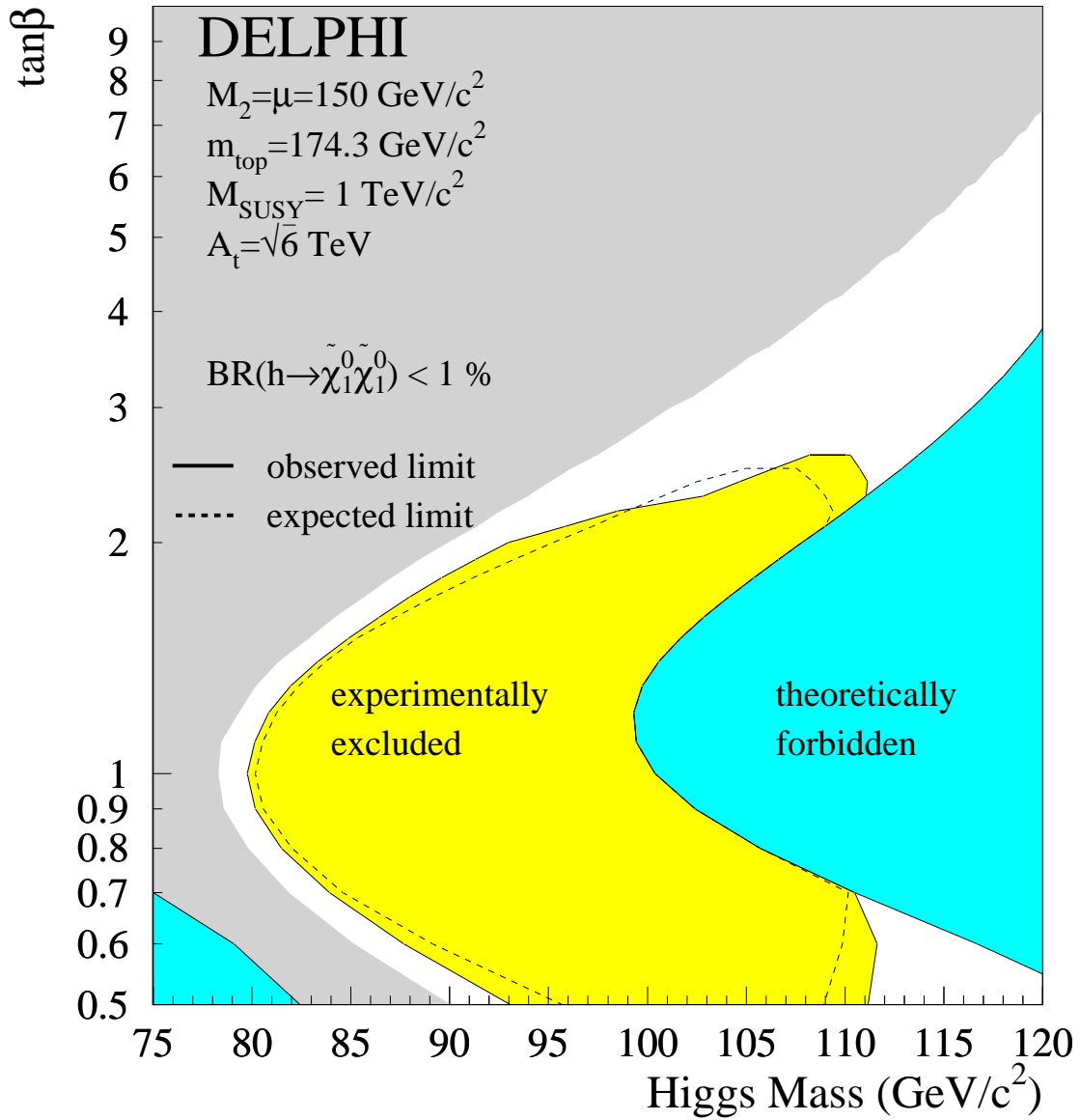}
\caption[]{
Excluded region in the MSSM from searches for a 
Higgs boson decaying into invisible final states
for the modified ``$\rm m_h$-max scenario'' described in the text.
%$\mathrm{m_h}$-max in the stop sector.
The different grey areas show the theoretically forbidden region (dark),
the region where the Higgs boson does not decay into neutralinos 
(intermediate),
the region which is excluded at 95\% CL by this search for invisibly 
decaying Higgs bosons (light) and the unexcluded region (white).}
\label{fig:mssmmax}
\end{center}
\end{figure}

In the MSSM, there are parameter regions where the Higgs boson can decay 
into neutralinos, $\tilde{\chi}^0$, which leads to invisible Higgs decays.
As an illustration a benchmark scenario including such decays
was defined from the so-called ``$\rm m_h$-max scenario''~\cite{DELPHIh2000}. 
In this scenario the MSSM parameters are the 
mass of the pseudoscalar Higgs boson, $m_{\rm A}$, the ratio of the vacuum
expectation values, $\tan\beta$, the mixing in the scalar top sector $X_{\rm t}$,
the gaugino mass $M_2$ and the Higgs self-coupling $\mu$.
$M_2$ and $\mu$ were modified to obtain light neutralino masses setting 
$M_2=\mu=150$~\GeVcc.
Then, a scan was performed in the $\tan\beta$-$m_{\rm A}$ plane.
For each scan point the hZ production cross-section and the Higgs boson
branching ratio into neutralinos were calculated, and the 
point was considered as excluded if the product 
%of cross-section for a Higgs boson decaying into neutralinos 
was found to be larger
than the excluded cross-section as shown in Fig.~\ref{pic:lim_inv}.
Figure~\ref{fig:mssmmax} shows the 
excluded region from the search for invisible Higgs decays,
the theoretically forbidden region, and 
the region where the branching ratio 
$\rm h\rightarrow \tilde{\chi}^0\tilde{\chi}^0$ is less than 1\%.
In this benchmark scenario, the invisible Higgs boson search
covers a large region in the low $\tan\beta$ regime.  
The white regions cannot be excluded by the invisible Higgs searches 
alone because the branching ratio into neutralinos is too small.
%The decay branching fractions depend on the model parameter combinations.
%Most of the unexcluded parameter region is excluded
%when in addition the decays into b quarks and $\tau$ leptons 
%are considered~\cite{DELPHIh2000}.
The search for the invisible Higgs boson decays also sets limits
in the general framework searches for Supersymmetric 
particles~\cite{delphisusy} and for searches in Anomaly Mediated Supersymmetry
Breaking (AMSB) models~\cite{amsb}.

\section{Conclusion \label{sec:CONCLU}}
In the data samples collected by the DELPHI detector at
centre-of-mass energies from 189 to 209~\GeV, 153 $\mathrm{q\bar{q}}$ 
(213 for the low mass analyses), 18 $\mathrm{\mu^+\mu^-}$, 
20 $\mathrm{e^+e^-}$ and 37 $\mathrm{\tau^+\tau^-}$ 
events were selected in searches for a Higgs boson decaying into 
invisible final states.
These numbers are consistent with the expectation from SM
background processes.

We set a 95\% CL lower mass limit of 112.1~\GeVcc\ for Higgs bosons 
with a Standard Model cross-section and with 100\% branching fraction 
into invisible decays.
%By combining this search for invisible decays with previous limits on 
%visible decays we set a 95\% CL lower mass limit of 112.0 \GeVcc\
%for a Higgs boson with an arbitrary invisible branching fraction.
Excluded parameter regions are given in a simple Majoron model. 
The invisible Higgs boson search is important to cover some parameter 
regions in the MSSM where Higgs decays into neutralinos are kinematically allowed.

%         Modified on 04-06-1999 by dimartino
%-------------------------------------------------------------------
\subsection*{Acknowledgements}
\vskip 3 mm
 We are greatly indebted to our technical 
collaborators, to the members of the CERN-SL Division for the excellent 
performance of the LEP collider, and to the funding agencies for their
support in building and operating the DELPHI detector.\\
We acknowledge in particular the support of \\
Austrian Federal Ministry of Education, Science and Culture,
GZ 616.364/2-III/2a/98, \\
FNRS--FWO, Flanders Institute to encourage scientific and technological 
research in the industry (IWT), Federal Office for Scientific, Technical
and Cultural affairs (OSTC), Belgium,  \\
FINEP, CNPq, CAPES, FUJB and FAPERJ, Brazil, \\
Czech Ministry of Industry and Trade, GA CR 202/99/1362,\\
Commission of the European Communities (DG XII), \\
Direction des Sciences de la Mati$\grave{\mbox{\rm e}}$re, CEA, France, \\
Bundesministerium f$\ddot{\mbox{\rm u}}$r Bildung, Wissenschaft, Forschung 
und Technologie, Germany,\\
General Secretariat for Research and Technology, Greece, \\
National Science Foundation (NWO) and Foundation for Research on Matter (FOM),
The Netherlands, \\
Norwegian Research Council,  \\
State Committee for Scientific Research, Poland, SPUB-M/CERN/PO3/DZ296/2000,
SPUB-M/CERN/PO3/DZ297/2000 and 2P03B 104 19 and 2P03B 69 23(2002-2004)\\
JNICT--Junta Nacional de Investiga\c{c}\~{a}o Cient\'{\i}fica 
e Tecnol$\acute{\mbox{\rm o}}$gica, Portugal, \\
Vedecka grantova agentura MS SR, Slovakia, Nr. 95/5195/134, \\
Ministry of Science and Technology of the Republic of Slovenia, \\
CICYT, Spain, AEN99-0950 and AEN99-0761,  \\
The Swedish Natural Science Research Council,      \\
Particle Physics and Astronomy Research Council, UK, \\
Department of Energy, USA, DE-FG02-01ER41155, \\
EEC RTN contract HPRN-CT-00292-2002. \\
%=========================================================================%

%=========================================================================%

\clearpage
\newpage

\end{document}